\documentclass[aps,prb,twocolumn]{revtex4-1}%

\usepackage{amsmath}
\usepackage{graphicx} 
\usepackage{bm} 
\usepackage{color} 
\usepackage{amssymb}
\usepackage{verbatim}
\usepackage[colorlinks=true, pdfstartview=FitV, linkcolor=blue, citecolor=blue, urlcolor=blue]{hyperref} 

\newcommand{\ee}[1]{\cdot10^{#1}}
\newcommand{\mr}[1]{\mathrm{#1}}
\newcommand{\unit}[1]{\,\mathrm{#1}}
\newcommand{\um}{\,\mu{\rm m}}
\newcommand{\us}{\,\mu{\rm s}}

\newcommand{\uA}{\,\mu{\rm A}}

\newcommand{\rtHz}{\sqrt{\mr{Hz}}}

\newcommand{\degree}{^\circ}

\newcommand{\ye}{\gamma}

\newcommand{\ex}{e_x}
\newcommand{\ey}{e_y}
\newcommand{\ez}{e_z}
\newcommand{\kx}{k_x}
\newcommand{\ky}{k_y}

\newcommand{\jx}{J_x}
\newcommand{\jy}{J_y}

\newcommand{\lmax}{\lambda}
\newcommand{\lcrit}{\lambda_\mr{crit}}
\newcommand{\heff}{h_\mr{eff}}
\newcommand{\vecn}{{\bf n}}

\newcommand{\vecBdc}{{\bf B}_\mr{dc}}
\newcommand{\vecBmw}{{\bf B}_\mr{mw}}
\newcommand{\vecJ}{{\bf J}}

\newcommand{\Bpar}{B_{||}}
\newcommand{\Bperp}{B_{\perp}}

\newcommand{\wo}{\omega_0}
\newcommand{\woIp}{\omega_{+I}}
\newcommand{\woIm}{\omega_{-I}}

\newcommand{\dw}{\delta\omega}
\newcommand{\Dw}{\Omega}

\newcommand{\eps}{\epsilon}
\newcommand{\epsmax}{\epsilon_\mr{max}}

\begin{document}

\title{Nanoscale imaging of current density with a single-spin magnetometer}
\author{K. Chang}
\author{A. Eichler}
\author{C. L. Degen}
\email{degenc@ethz.ch}
\affiliation{Department of Physics, ETH Zurich, Otto Stern Weg 1, 8093 Zurich, Switzerland}
\date{\today}
\begin{abstract}
Charge transport in nanostructures and thin films is fundamental to many phenomena and processes in science and technology, ranging from quantum effects and electronic correlations in mesoscopic physics, to integrated charge- or spin-based electronic circuits, to photoactive layers in energy research.  Direct visualization of the charge flow in such structures is challenging due to their nanometer size and the itinerant nature of currents.
In this work, we demonstrate non-invasive magnetic imaging of current density in two-dimensional conductor networks including metallic nanowires and carbon nanotubes.  Our sensor is the electronic spin of a diamond nitrogen-vacancy center attached to a scanning tip.  Using a differential measurement technique, we detect DC currents down to a few $\unit{\mu A}$ above a baseline current density of $\sim 2\ee{4}\unit{A/cm^2}$.  Reconstructed images have a spatial resolution of typically 50 nm, with a best-effort value of 22 nm.  Current density imaging offers a new route for studying electronic transport and conductance variations in two-dimensional materials and devices, with many exciting applications in condensed matter physics.
\end{abstract}
%


\maketitle

\section{Introduction}

Non-invasive detection of currents is possible thanks to the long-range magnetic field that appears near moving charges, according to the law of Biot and Savart \cite{jackson75}.  Although a map of the Oersted field does not directly reproduce an image of current flow, the current density can for some geometries be rigorously reconstructed.  A particularly important class of conductors are two-dimensional structures, such as patterned electronic circuits, semiconductor electron and hole gases, or organic and inorganic thin films.  In this case it is possible to reconstruct the two-dimensional current density from a single component of the magnetic field, recorded in a plane at a fixed distance over the conductor \cite{roth89}.  Millimeter-to-micrometer current density mapping has been performed by scanning Hall probes \cite{xing94}, magneto-optical methods \cite{johansen96}, magnetoresistance probes \cite{schrag03}, scanning SQUIDs \cite{knauss01,nowack13,shibata15}, and diamond chips with large ensembles of nitrogen-vacancy (NV) centers \cite{steinert10,chipaux15,nowodzinski15,tetienne16}.  

Despite of many challenges, there is a strong incentive to extend current density mapping to higher resolution, especially if the nanometer regime can be reached.  At the nanoscale, many interesting phenomena may be explored, such as branched electron flow \cite{topinka00,topinka01}, weak localization and universal conductance fluctuations \cite{ihn10}, edge conductance \cite{son06,zarbo07}, dissipation-less currents \cite{bleszynski09,bluhm09current}, or impurity back-scattering \cite{friedel52,cheianov06}.  A nanometer scale imaging capability could hence play an important role in mesoscopic condensed matter physics and guide the development of novel materials and circuits.  Impressive advances in sub-micrometer current density mapping have recently been made with nano-SQUIDs fabricated on the ends of pulled glass capillaries \cite{anahory14,zeldov16}.

In this work, we demonstrate sub-30-nm-resolution imaging of current density in patterned nanowire devices at room temperature using a scanning diamond magnetometer.  We find that the technique is ideally suited for current density mapping, as the scanning sensor spin is point-like and naturally provides a single component of the vector magnetic field.  We further show that DC and microwave currents can be mapped and reconstructed separately, providing two independent means for analyzing local current flow.  Finally, we demonstrate the potential of the technique by imaging currents flowing in a bundle of carbon nanotubes.

Diamond magnetometry uses a single defect spin in a diamond tip to sense the local magnetic field near a sample \cite{degen08apl}.  Optically-detected electron paramagnetic resonance (EPR) is used to probe the resonance of the spin via application of a microwave field and fluorescence detection \cite{gruber97,jelezko06}.  Because the spin resonance shifts with DC magnetic field via the Zeeman effect, the resonance frequency is directly proportional to the local magnetic field.  By controlled scanning of the diamond tip over the sample, high resolution magnetic images can be obtained.  Recent work has mostly focused on magnetic nanostructures \cite{balasubramanian08} including single electron spins \cite{grinolds13}, thin magnetic films \cite{rondin12,maletinsky12,rondin13,tetienne15} and surfaces of superconductors \cite{pelliccione16,thiel16}.
In the present work we apply the technique to image charge currents via the Oersted magnetic field.

\section{Apparatus}

\begin{figure}
\includegraphics[width=1.00\columnwidth]{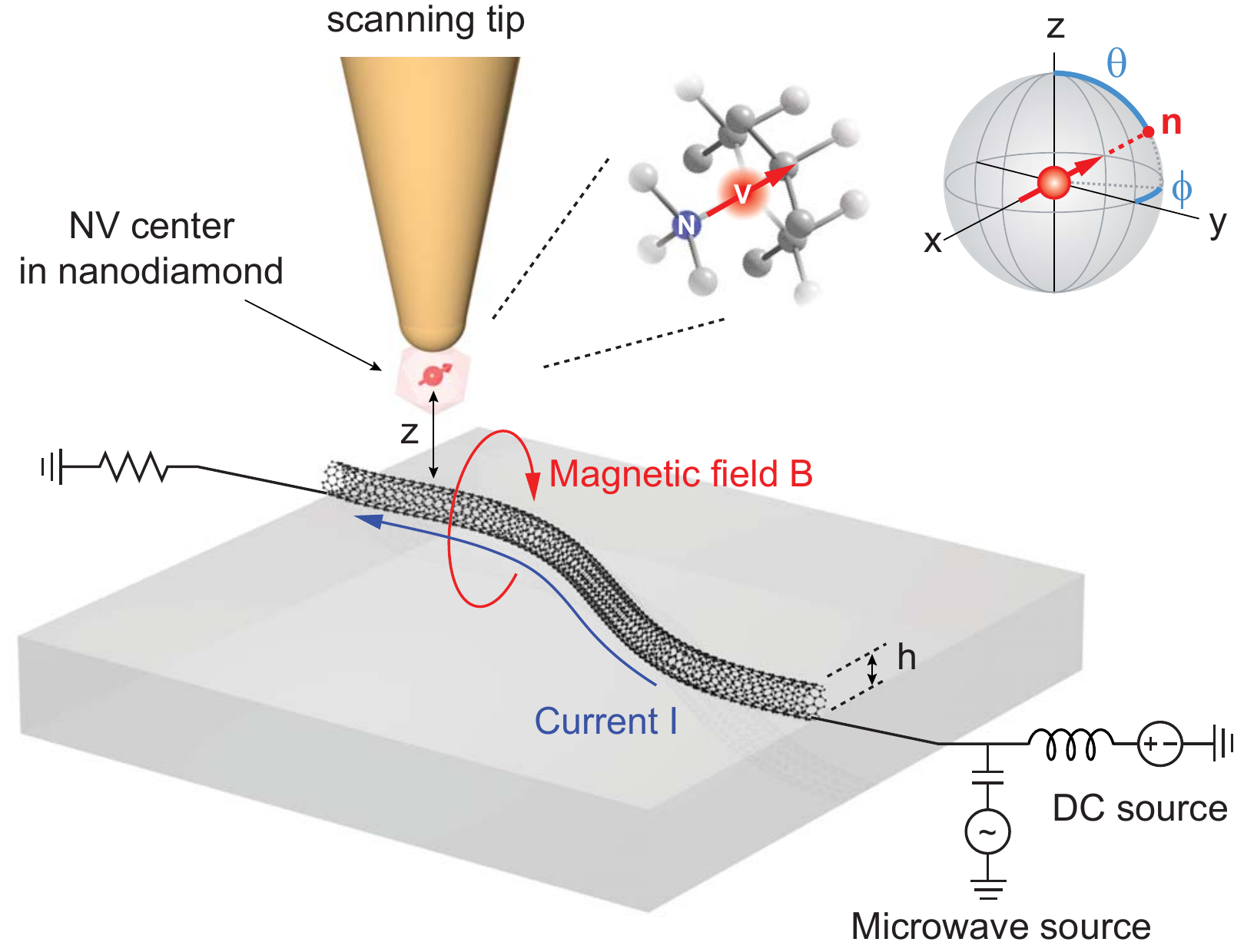}
\caption{
{Setup and components of the scanning diamond magnetometer.}
A scanning probe with a nitrogen-vacancy (NV) center at the end is positioned $z<100\unit{nm}$ over a current-carrying nanostructure.  The electron spin resonance (EPR) frequency of the NV center is continuously measured using microwave irradiation and optical fluorescence detection with a nearby objective (not shown).  The magnetic field generated by the DC current causes a Zeeman shift of the EPR frequency that can be converted to units of magnetic field.  The NV center only responds to fields that are parallel to its symmetry axis $\vecn$, defined by the angles $\theta$ and $\phi$ (see inset).
To map the current density, the magnetic field is recorded in an $(x,y)$ plane at a fixed distance $z$ from the surface followed by an image reconstruction.
Experiments are carried under ambient conditions and in a small static bias field of $\sim 4\unit{mT}$. 
}
\label{fig:fig1}
\end{figure}

The key element in our measurement apparatus is a scanning tip with a single NV center at its apex (Fig. \ref{fig:fig1}).  Peripheral instrumentation permits optical pumping and readout of the NV spin and its manipulation by microwave magnetic fields.  The tips are prepared by picking up $\sim 25\unit{nm}$-diameter diamond nanoparticles with a commercial atomic force microscopy (AFM) cantilever \cite{kuhn01,balasubramanian08,rondin12}. The symmetry axis of attached NV centers, which defines the ($\theta,\phi$) vector orientation of the sensor (see Fig. \ref{fig:fig1}), is determined via EPR spectroscopy.  The distance $z$ between the NV center and the sample surface can be inferred from one-dimensional line scans across a current-carrying nanowire (see below).

\begin{figure}
\includegraphics[width=1.00\columnwidth]{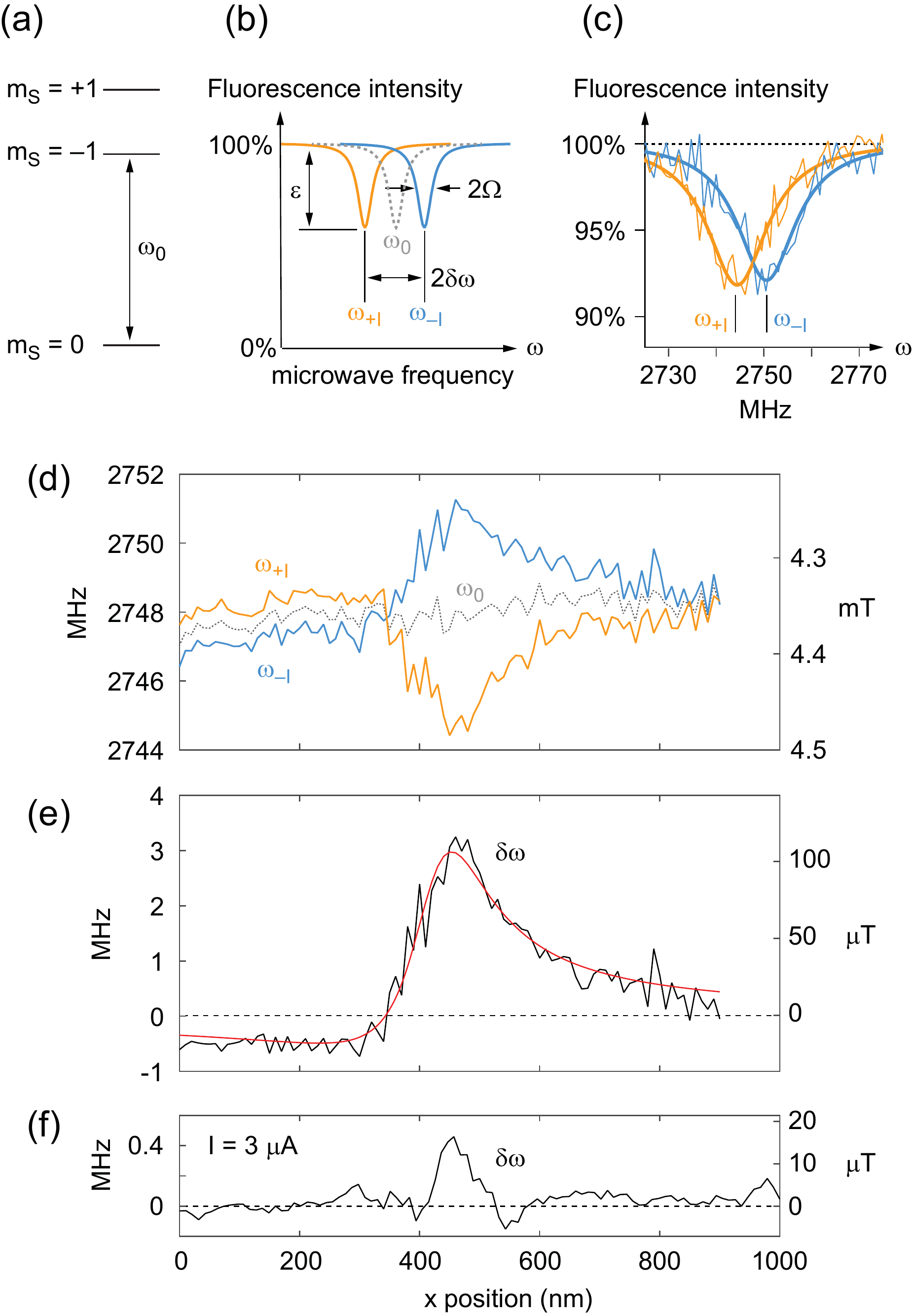}
\caption{
{Basic protocol for current measurements.}
{(a)} Energy level diagram of the NV center indicating the $m_s=0$ to $m_s=-1$ sensing EPR transition.
{(b)} Differential measurement scheme:  Two EPR spectra are recorded with positive or negative DC current $\pm I$ applied, leading to opposite shifts of the peak frequencies $\woIp$ and $\woIm$.  $\dw$ is the differential frequency shift, $\wo=\tfrac12(\woIp+\woIm)$ is the center frequency in absence of any current, $\Dw$ the resonance line width, and $\eps$ the optical contrast.
{(c)} Example experimental data illustrating the schematic in (b).
{(d)} Line scan across a $100\times100\unit{nm^2}$ Pt nanowire centered at $x=420\unit{nm}$ with an applied current of $I=96\unit{\uA}$.
Right scale converts the frequencies to units of magnetic field.
{(e)} Differential line shift $\dw$.  The bold red line represents a fit to the Oersted magnetic field expected from an infinite straight wire with stand-off $z=27\unit{nm}$. The NV center was oriented along ($\theta,\phi$) = ($70^\circ,67^\circ$).
{(f)} Line scan acquired with a different NV tip and a current of $3\unit{\uA}$.
}
\label{fig:fig2}
\end{figure}

The magnetic field is measured by recording a continuous-wave EPR spectrum and determining the peak frequency of the resonance using a Lorentzian fit \cite{dreau11,supplemental} (see Fig. \ref{fig:fig2}a-c).  To eliminate long-term electrical or thermal drift and to discriminate the current-induced field from other magnetic fields, we have implemented a differential measurement technique.  At each measurement location, two EPR spectra are recorded with a positive ($+I$) or negative ($-I$) source current applied to the device.  Spectra are taken by square wave modulation of the DC source at 1 kHz and binning photon counts in synchrony with the modulation.  In this way, low-frequency drift and background signals are efficiently rejected.  The difference between the EPR peak frequencies of the two spectra, $\dw = \frac12(\woIp-\woIm)$ [see Fig. \ref{fig:fig2}b], then directly corresponds to the component of the Oersted field that is parallel to the NV axis,
\begin{equation}
\Bpar = \vecBdc \cdot \vecn = \frac{\dw}{\ye} = \frac{\woIp-\woIm}{2\ye} \ ,
\label{eq:bpar}
\end{equation}
where $\ye = 28\unit{GHz/T}$ is the gyromagnetic ratio of the electronic spin and ${\bf n} = (\sin\theta\cos\phi, \sin\theta\sin\phi, \cos\theta)$ is the unit vector along the NV symmetry axis (see Fig. \ref{fig:fig1}).
The Lorentzian fit also yields values for the line width parameter $\Dw$ and the optical contrast $\eps$ (see Fig. \ref{fig:fig2}b) which can be used to infer the local microwave field $\vecBmw$,
\begin{equation}
\Bperp = |\vecBmw\times\vecn| \approx \frac{2\Dw}{\ye}\sqrt{\frac{\eps}{\epsmax}}
\label{eq:bperp}
\end{equation}
where $\epsmax$ is the saturated optical contrast \cite{dreau11,supplemental}.

To develop our current imaging technique, we fabricate several sets of nanowire test devices with different geometries, including straight sections, turns, kinks and splits.  The metallic nanowires are made by e-beam lithography and Pt deposition and have cross-sections between $50\times50$ and $100\times100 \unit{nm^2}$ (see Ref. \onlinecite{supplemental}).  The devices can be connected to both a DC and microwave source ($\sim2.7\unit{GHz}$) such that the same nanowire can be used for producing DC fields and actuating the EPR transition.
To test a device, we scan the NV center laterally across a straight section of a nanowire while applying a current, and record the differential line shift $\dw$ as a function of $x$ position (Fig. \ref{fig:fig2}d,e).  This line scan can be compared to an analytical model and allows us to accurately calibrate the stand-off distance $z$; for the tips used in this study, stand-offs were typically between $25\unit{nm}$ and $100\unit{nm}$.

The line scans also provide an estimate of the minimum detectable current.  For DC currents, this value is around $\sim 1\unit{\uA}$ (Fig. \ref{fig:fig2}f).
The minimum detectable microwave current is in principle much lower, and ultimately limited by the $T_1$ time of the NV spin \cite{appel15}.  For $T_1 = 10\unit{\us}$, the magnetic field noise is $\sim 2.5\unit{nT/\rtHz}$, corresponding to a noise in the microwave current of a few $\mr{nA/\rtHz}$.  However, because microwave imaging requires precise matching of the spin energy levels to the microwave frequency, we found it difficult to detect microwave currents $<50\unit{\uA}$ (Ref. \onlinecite{supplemental}).

\section{Magnetic imaging}

\begin{figure*}
\includegraphics[width=0.9\textwidth]{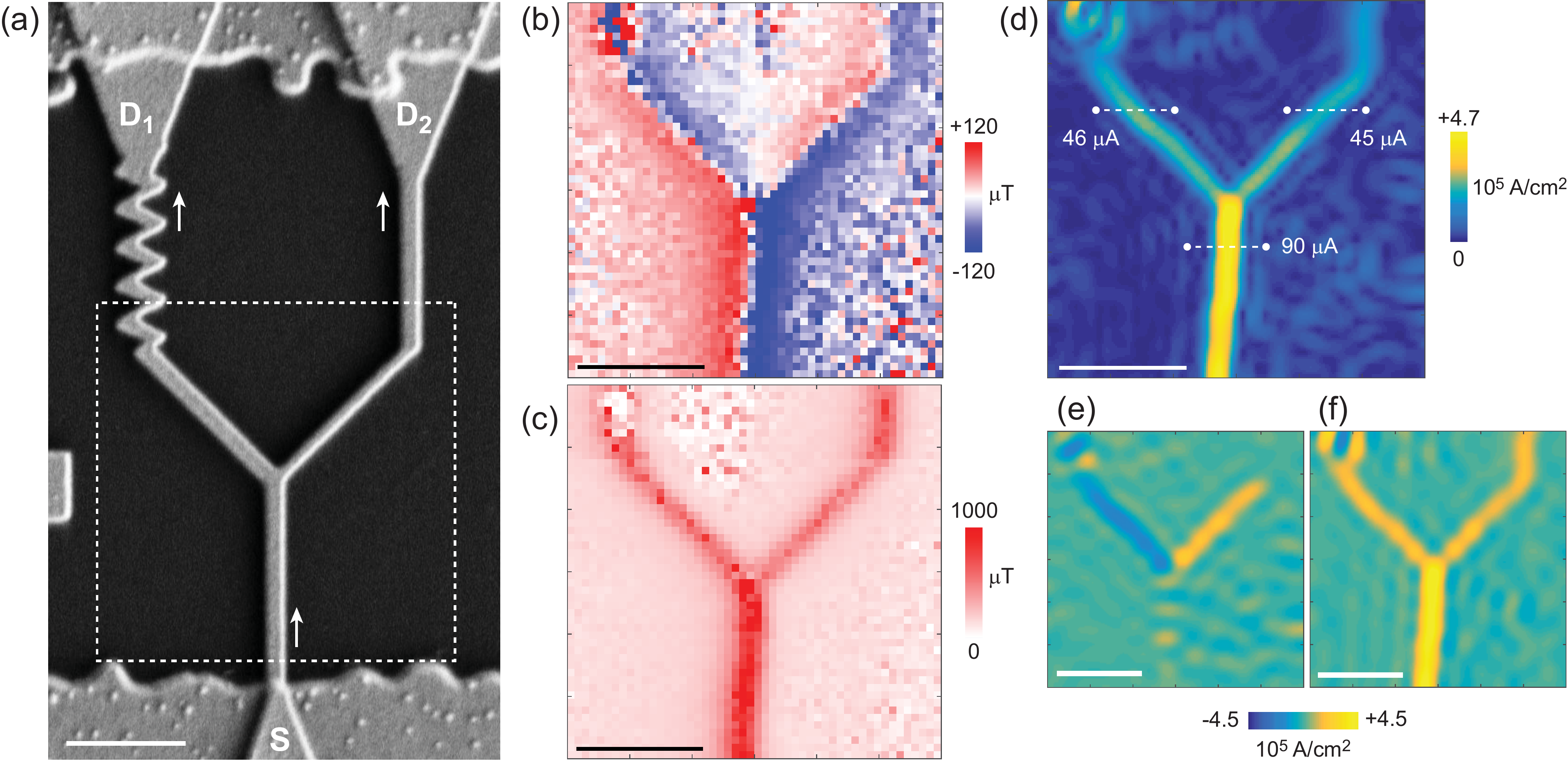}
\caption{
{Two-dimensional images of geometry, magnetic field and reconstructed current density for a Pt nanowire test device.}
{(a)}, Scanning electron micrograph of the device indicating the direction of current flow ($I=90\unit{\uA}$).
{(b)}, Image (raw data) of the differential line shift, representing the component $\Bpar$ of the DC magnetic field that is parallel to the NV center axis $\vecn$.  The scanning NV center had a stand off $z\approx 75\unit{nm}$ and an orientation $(\theta,\phi)=(2\degree,322\degree$).  Images are composed of an array of $50\times50$ pixels spaced by $60\unit{nm}$, where each pixel required one minute of averaging.
The pixelated areas correspond to regions where the EPR spectrum was not well resolved and the Lorentzian fit error was large.
{(c)}, Image (raw data) of the line width, representing the component $\Bperp$ of the microwave magnetic field that is transverse to $\vecn$.
{(d)}, Current density image $|{\bf J}(x,y)|$ reconstructed from $\Bpar$ using a spatial filtering parameter of $\lmax = 215\unit{nm}$.
The corresponding reconstruction from $\Bperp$ is provided in Supplemental Fig. S2.
{(e)}, $\jx$ and $\jy$ components of the current density.
All scale bars are $1\unit{\um}$.
}
\label{fig:fig3}
\end{figure*}

To demonstrate two-dimensional imaging, we use a Y-shaped structure where the current is injected at the bottom (S) and collected by the two arms at the top (D$_1$ and D$_2$, see electron micrograph in Fig. \ref{fig:fig3}a).  To form a magnetic image of the current flowing in this device, the diamond tip is scanned laterally in a plane at fixed $z$ spacing from the top of the wire (Fig. \ref{fig:fig1}).  The spacing is maintained by briefly approaching and retracting the tip by a known amount before each point.  The resulting magnetic image, shown in Fig. \ref{fig:fig3}b, clearly reflects the geometry of the underlying structure.  Since this image is taken with an NV center oriented nearly parallel to the $z$ axis, the magnetic field is positive to one side of the wire while it is negative to the other side.

In addition to the DC magnetic field, we can also image the microwave magnetic field by plotting the line width parameter $\Dw$ (Fig. \ref{fig:fig3}c).  For sufficiently strong microwave fields where the optical contrast is saturated ($\eps \approx \epsmax$), microwave amplitude and line width are directly proportional, $\Bperp \approx 2\Dw/\ye$ [see Eq. (\ref{eq:bperp})].  Because both the DC and microwave fields originate from the same conductor (and neglecting the effects of skin depth and wavelength), it can be expected that the two fields have the same spatial distribution.
The vector component picked up in the two imaging modes is, however, orthogonal.  While the DC image corresponds to the magnetic field parallel to the spin orientation $\vecn$, the microwave image represents the component transverse to $\vecn$.  This feature can be observed in Fig. \ref{fig:fig3}b,c, where the DC field passes through zero right above the conductor while the microwave field is maximized at this location, as expected from the approximately vertical orientation of this NV center.

\section{Reconstruction of current density}

In a next step, we reconstruct the two-dimensional current density ${\bf J}(x,y) = (\jx,\jy)$ from the DC magnetic image $\Bpar(x,y)$.  This can be achieved by inverting Biot and Savart's law.  For this purpose we adapt an inverse filtering technique described by Roth et al. \onlinecite{roth89}.  As discussed in the Supplemental Material (Ref. \cite{supplemental}), the reconstructed current density in Fourier space is
\begin{eqnarray}
\jx(\kx,\ky) & = & \frac{w(k,\lmax) \ky}{g(k,z)\left[\ey\ky - \ex\kx + i\ez k\right]} \Bpar(\kx,\ky,z) \  \\
\jy(\kx,\ky) & = & \frac{w(k,\lmax) \kx}{g(k,z)\left[\ex\kx - \ey\ky - i\ez k\right]} \Bpar(\kx,\ky,z) \ 
\end{eqnarray}
where $\kx,\ky$ are $k$-space vectors, $k = (\kx^2+\ky^2)^{1/2}$ and where $(\ex,\ey,\ez)={\bf n}$ is the sensor orientation.  A similar set of equations can be derived to infer $\vecJ(x,y)$ from the microwave field $\Bperp(x,y)$ \cite{supplemental}.  The function $g(k,z)$ is the Green's function
\begin{equation}
g(k,z) = \frac{\mu_0 \heff}{2} e^{-k z} \ ,
\label{eq:g}
\end{equation}
where $\heff = (1-e^{-k h})/k$ is an effective thickness and $h$ the physical thickness of the conductor (see Fig. \ref{fig:fig1}).  In addition, $w(k,\lmax)$ is a window function required to suppress noise at high spatial frequencies $k$ where $g(k,z)$ is small.  We use a Hanning window,
\begin{equation}
w(k,\lmax) =
\left\{
\begin{array}{cl}
\frac12\left[1+\cos(\tfrac12 k\lmax)\right] & \text{if}\ |k| < 2\pi/\lmax \\
0 & \text{otherwise}
\end{array}
\right. \ ,
\label{eq:w}
\end{equation}
with cut-off wavelength $\lmax$.  The wavelength $\lmax$ plays the role of a spatial filtering parameter that can be tuned to adjust image resolution and noise rejection (see below).  Beside the Hanning filter, several optional image processing steps are found to significantly improve the quality of the reconstruction \cite{supplemental}.

Figs. \ref{fig:fig3}d-f display the results of the reconstruction, which include the magnitude of the current density $|\vecJ|$ as well as the individual components $\jx$ and $\jy$.  Given that the raw data is rather coarsely sampled and has limited signal-to-noise ratio (SNR), the quality of the reconstruction is quite remarkable.  This apparent improvement is expected, because the reconstruction process inherently acts as a spatial low-pass filter through the cut-off parameter $\lmax$.  Excellent agreement is found with the scanning electron micrograph of Fig. \ref{fig:fig3}a.  Even fine features, such as the zig-zag structure at the top left corner, are reasonably well resolved, and positive and negative current densities are reliably reproduced.  The magnitudes of the currents entering and exiting through the three arms quantitatively agree with Kirchhoff's rule and the nominal source current.

\begin{figure*}
\includegraphics[width=0.85\textwidth]{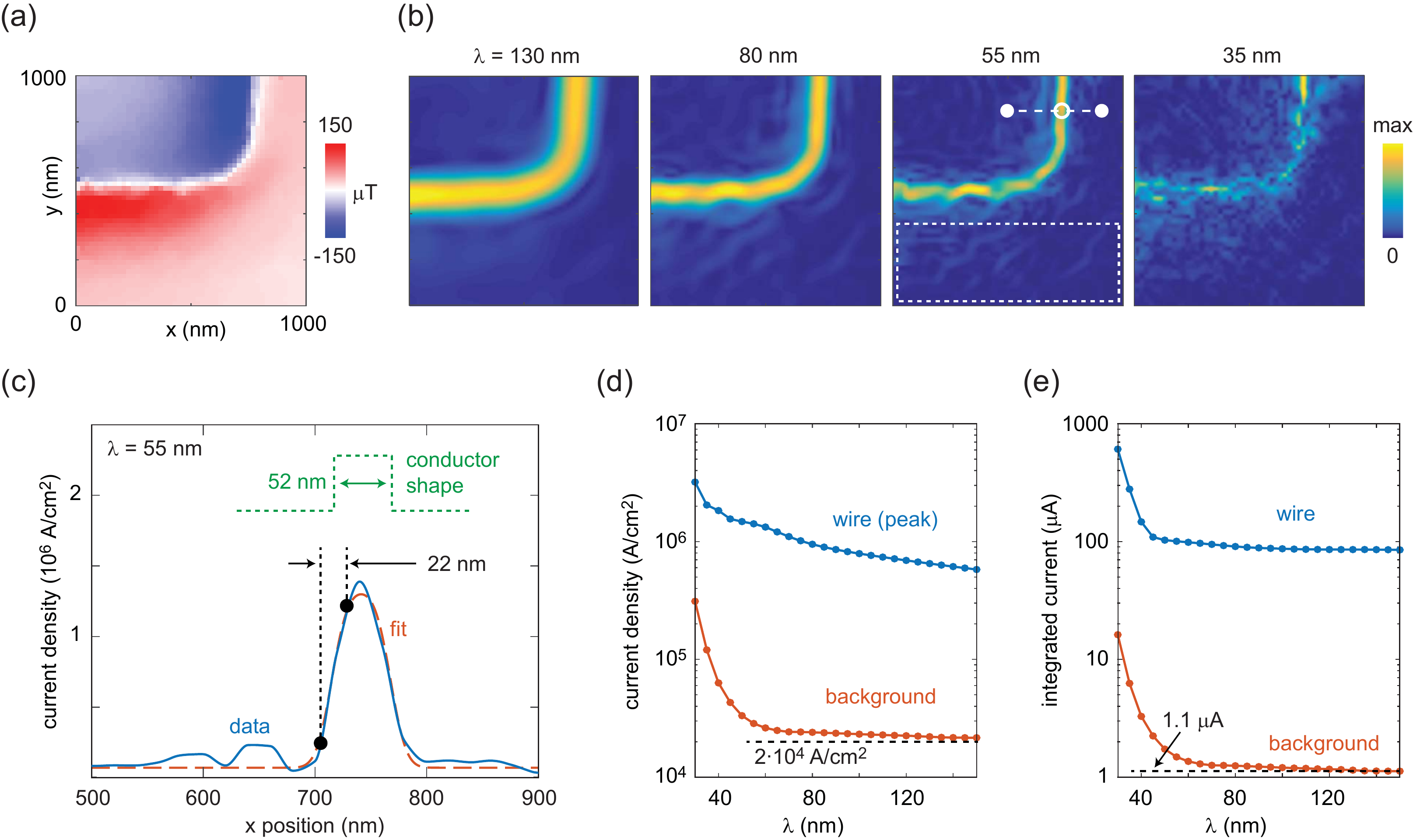}
\caption{
{Demonstration of 22 nm spatial resolution.}
{(a)}, Magnetic image of an elbow-shaped Pt nanowire with a source current of $96\unit{\uA}$.
{(b)}, Current density $|{\bf J}|$ reconstructed for decreasing values of the spatial filter parameter $\lmax$.
{(c)}, One-dimensional plot along the dashed line in (b).  The blue solid line represents the data and the red dashed curve represents the conductor shape (green) convolved with a Gaussian.  The spatial resolution, indicated by the 15\%-85\% rise of the signal, is $\sim 22\unit{nm}$.
{(d)}, Root-mean-square noise of the background (dotted rectangle in b) and peak current density of the wire (hollow dot in b) as a function of $\lmax$.
{(e)}, Integrated current and corresponding background current as a function of $\lmax$.  Integration of nanowire current was done over dashed line in (b) and over the nanowire height ($h \sim 100\unit{nm}$).  Background current represents background current density multiplied by conductor cross-section.
}
\label{fig:fig4}
\end{figure*}
%

\section{Spatial resolution}

A key feature of scanning diamond magnetometry is the technique's potential for imaging with high spatial resolution and sensitivity.  In our experiment, the resolution can be tuned through the filter parameter $\lmax$.  By reducing $\lmax$, the resolution is refined up to the point where the noise in the reconstructed image becomes excessive (see Fig. \ref{fig:fig4}a,b).  This occurs at a certain critical wavelength $\lcrit$ that is of order of the tip stand-off $z$, here $\lcrit \approx 50\unit{nm}$.  The critical wavelength is due to the exponential factor in the Green's function $g\propto e^{-kz}$ (Eq. \ref{eq:g}) and the maximum allowed wave vector $k=2\pi/\lmax$ (Eq. \ref{eq:w}).
To estimate the image resolution, we can inspect a line cut across the conductor for $\lmax \approx \lcrit$ (Fig. \ref{fig:fig4}c).
The spatial resolution, defined as the $x$ distance over which the signal rises from 15\% to 85\% of its maximum, is about $22\unit{nm}$.
This distance corresponds to $2\sigma$ of a Gaussian convolved with the conductor shape (Sparrow's criterion \cite{sparrow16,jones95}),
and is consistent with the fastest permitted spatial oscillation of $\lmax/4 \sim 14 \unit{nm} \leq 2\sigma$.

We can further determine the baseline current density and integrated conductor current as a function of $\lmax$ (Figs. \ref{fig:fig4}d,e).  We find that both quantities are approximately constant when $\lmax$ is chosen larger than the critical wavelength $\lcrit$.  For the dataset in Fig. \ref{fig:fig4}, the baseline current density (root-mean-square value) is of order $2\ee{4}\unit{A/cm^2}$.  This converts to a minimum detectable current of $\sim 1\unit{\uA}$ (Fig. \ref{fig:fig4}e), in agreement with our earlier finding.

%
\begin{figure*}
\includegraphics[width=0.9\textwidth]{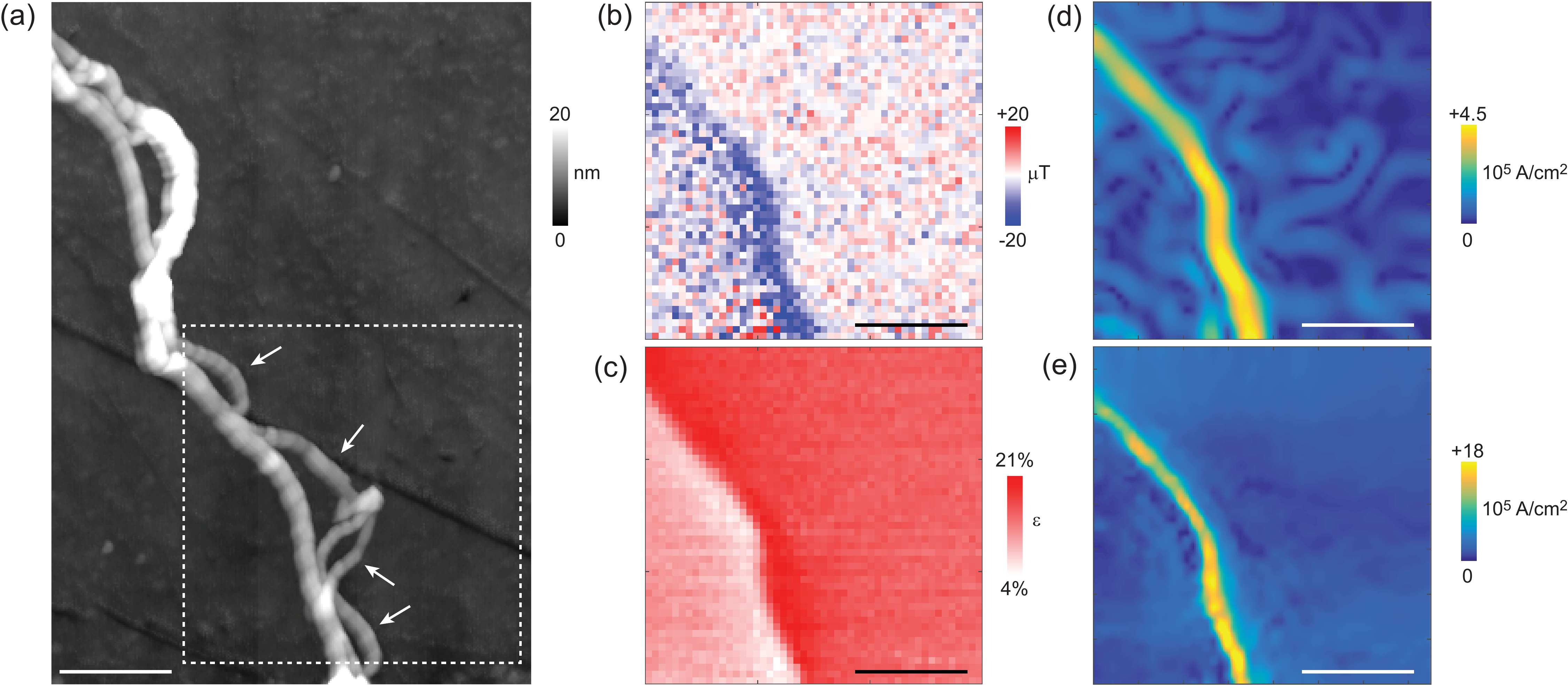}
\caption{
{Imaging results for carbon nanotubes (CNTs).}
{(a)}, Topographic image recorded using an atomic force microscope.  The typical CNT height is 10 nm.
{(b)}, Image (raw data) of the differential line shift $\dw/\ye$, representing the DC magnetic field $\Bpar$.
{(c)}, Image (raw data) of the optical contrast $\eps$, representing the microwave magnetic field $\Bperp$. $\epsmax$ was 0.23.
{(d,e)}, Current density $|{\bf J}|$ reconstructed from (b) and (c), respectively.  Clearly, only the center tube is conducting and none of the extra loops (arrows in a) carry any current.  Nominal source current was $10\unit{\uA}$, tip stand-off was $40\unit{nm}$, and ($\theta,\phi$) = ($66\degree,220\degree$).  Filter parameters were $\lmax = 130\unit{nm}$ and $60\unit{nm}$ for (d) and (e), respectively.  Scale bars are $500\unit{nm}$.
}
\label{fig:fig5}
\end{figure*}
%

\section{Carbon nanotube imaging}

Finally, we explore the application of the technique to detect currents flowing in bundles of carbon nanotubes (CNTs).  The CNTs were vertically grown on Si  with Fe as the catalyst \cite{Youn13}, dispersed onto a quartz substrate, and contacted by e-beam lithography \cite{supplemental}.  Fig. \ref{fig:fig5} shows an example of CNT imaging.  In this experiment, the line shift $\dw$ has a relatively low signal-to-noise ratio (Fig. \ref{fig:fig5}b) and the optical contrast $\eps$ (Fig. \ref{fig:fig5}c) provides the strongest image contrast.  The line width $\Dw\approx 3\unit{MHz}$ is approximately constant (Supplemental Fig. S3).  This represents the regime of low microwave amplitude where the EPR transition is not saturated and $\eps\propto\sqrt{\Bperp}$ (see Eq. (\ref{eq:bperp}) and Refs. \onlinecite{dreau11,supplemental}).  Both the $\dw$ and $\eps$ images can be reconstructed (Figs. \ref{fig:fig5}d,e), with significantly better resolution resulting from the optical contrast data.  The current density image reveals interesting features about the CNT bundle; in particular, only a center portion of the bundle appears to be conducting while a number of side loops (arrows in Fig. \ref{fig:fig5}a) do not carry any measurable current.

\section{Conclusions and outlook}

In summary, we have shown that the technique of scanning diamond magnetometry enables two-dimensional imaging of current density with sub-30-nm spatial resolution and $\sim 1\unit{\uA}$ sensitivity.  The magnetic imaging process is passive, with no direct disturbance of the electron flow, and can be carried out under ambient conditions.  Further improvements can be expected as the diamond probe and acquisition techniques are being refined.  For example, by replacing our nanodiamond probes with etched single-crystal diamond tips \cite{maletinsky12,pelliccione16,thiel16,appel16}, a resolution below $10\unit{nm}$ should be reachable.  Moreover, pulsed EPR techniques could be employed to detect signals via spin echoes \cite{taylor08,maze08} or optimized phase-estimation protocols \cite{nusran12,waldherr12,bonato16}, with expected magnetic sensitivities in the $10-100\unit{nT}$ range.  Together, these advances would lower the current sensitivity to near $1\unit{nA}$.  Such a capability would allow resolving even weak current density fluctuations, and possibly provide new insight into the local conductance of two-dimensional materials in condensed matter physics.
 
\section{Acknowledgments}

This work has been supported by the DIADEMS program 611143 by the European Commission, and by the Swiss NSF through the NCCR QSIT.
We thank J. Rhensius, A. Baumgartner, S. K. Youn and H. G. Park for aiding the CNT device fabrication and A. Dussaux, L. Lorenzelli, and J. Zopes for discussions and support.



\begin{thebibliography}{47}%
\makeatletter
\providecommand \@ifxundefined [1]{%
 \@ifx{#1\undefined}
}%
\providecommand \@ifnum [1]{%
 \ifnum #1\expandafter \@firstoftwo
 \else \expandafter \@secondoftwo
 \fi
}%
\providecommand \@ifx [1]{%
 \ifx #1\expandafter \@firstoftwo
 \else \expandafter \@secondoftwo
 \fi
}%
\providecommand \natexlab [1]{#1}%
\providecommand \enquote  [1]{``#1''}%
\providecommand \bibnamefont  [1]{#1}%
\providecommand \bibfnamefont [1]{#1}%
\providecommand \citenamefont [1]{#1}%
\providecommand \href@noop [0]{\@secondoftwo}%
\providecommand \href [0]{\begingroup \@sanitize@url \@href}%
\providecommand \@href[1]{\@@startlink{#1}\@@href}%
\providecommand \@@href[1]{\endgroup#1\@@endlink}%
\providecommand \@sanitize@url [0]{\catcode `\\12\catcode `\$12\catcode
  `\&12\catcode `\#12\catcode `\^12\catcode `\_12\catcode `\%12\relax}%
\providecommand \@@startlink[1]{}%
\providecommand \@@endlink[0]{}%
\providecommand \url  [0]{\begingroup\@sanitize@url \@url }%
\providecommand \@url [1]{\endgroup\@href {#1}{\urlprefix }}%
\providecommand \urlprefix  [0]{URL }%
\providecommand \Eprint [0]{\href }%
\providecommand \doibase [0]{http://dx.doi.org/}%
\providecommand \selectlanguage [0]{\@gobble}%
\providecommand \bibinfo  [0]{\@secondoftwo}%
\providecommand \bibfield  [0]{\@secondoftwo}%
\providecommand \translation [1]{[#1]}%
\providecommand \BibitemOpen [0]{}%
\providecommand \bibitemStop [0]{}%
\providecommand \bibitemNoStop [0]{.\EOS\space}%
\providecommand \EOS [0]{\spacefactor3000\relax}%
\providecommand \BibitemShut  [1]{\csname bibitem#1\endcsname}%
\let\auto@bib@innerbib\@empty
\bibitem [{\citenamefont {Jackson}(1975)}]{jackson75}%
  \BibitemOpen
  \bibfield  {author} {\bibinfo {author} {\bibfnamefont {J.}~\bibnamefont
  {Jackson}},\ }\href@noop {} {\emph {\bibinfo {title} {Classical
  electrodynamics}}}\ (\bibinfo  {publisher} {Wiley},\ \bibinfo {year}
  {1975})\BibitemShut {NoStop}%
\bibitem [{\citenamefont {Roth}\ \emph {et~al.}(1989)\citenamefont {Roth},
  \citenamefont {Sepulveda},\ and\ \citenamefont {Wikswo}}]{roth89}%
  \BibitemOpen
  \bibfield  {author} {\bibinfo {author} {\bibfnamefont {B.~J.}\ \bibnamefont
  {Roth}}, \bibinfo {author} {\bibfnamefont {N.~G.}\ \bibnamefont {Sepulveda}},
  \ and\ \bibinfo {author} {\bibfnamefont {J.~P.}\ \bibnamefont {Wikswo}},\
  }\href {\doibase 10.1063/1.342549} {\bibfield  {journal} {\bibinfo  {journal}
  {J. Appl. Phys.}\ }\textbf {\bibinfo {volume} {65}},\ \bibinfo {pages} {361}
  (\bibinfo {year} {1989})}\BibitemShut {NoStop}%
\bibitem [{\citenamefont {Xing}\ \emph {et~al.}(1994)\citenamefont {Xing},
  \citenamefont {Heinrich}, \citenamefont {Zhou}, \citenamefont {Fife},\ and\
  \citenamefont {Cragg}}]{xing94}%
  \BibitemOpen
  \bibfield  {author} {\bibinfo {author} {\bibfnamefont {W.}~\bibnamefont
  {Xing}}, \bibinfo {author} {\bibfnamefont {B.}~\bibnamefont {Heinrich}},
  \bibinfo {author} {\bibfnamefont {H.}~\bibnamefont {Zhou}}, \bibinfo {author}
  {\bibfnamefont {A.~A.}\ \bibnamefont {Fife}}, \ and\ \bibinfo {author}
  {\bibfnamefont {A.~R.}\ \bibnamefont {Cragg}},\ }\href {\doibase
  10.1063/1.357308} {\bibfield  {journal} {\bibinfo  {journal} {J. Appl. Phys}\
  }\textbf {\bibinfo {volume} {76}},\ \bibinfo {pages} {4244} (\bibinfo {year}
  {1994})}\BibitemShut {NoStop}%
\bibitem [{\citenamefont {Johansen}\ \emph {et~al.}(1996)\citenamefont
  {Johansen}, \citenamefont {Baziljevich}, \citenamefont {Bratsberg},
  \citenamefont {Galperin}, \citenamefont {Lindelof}, \citenamefont {Shen},\
  and\ \citenamefont {Vase}}]{johansen96}%
  \BibitemOpen
  \bibfield  {author} {\bibinfo {author} {\bibfnamefont {T.~H.}\ \bibnamefont
  {Johansen}}, \bibinfo {author} {\bibfnamefont {M.}~\bibnamefont
  {Baziljevich}}, \bibinfo {author} {\bibfnamefont {H.}~\bibnamefont
  {Bratsberg}}, \bibinfo {author} {\bibfnamefont {Y.}~\bibnamefont {Galperin}},
  \bibinfo {author} {\bibfnamefont {P.~E.}\ \bibnamefont {Lindelof}}, \bibinfo
  {author} {\bibfnamefont {Y.}~\bibnamefont {Shen}}, \ and\ \bibinfo {author}
  {\bibfnamefont {P.}~\bibnamefont {Vase}},\ }\href {\doibase
  10.1103/PhysRevB.54.16264} {\bibfield  {journal} {\bibinfo  {journal}
  {Physical Review B}\ }\textbf {\bibinfo {volume} {54}},\ \bibinfo {pages}
  {16264} (\bibinfo {year} {1996})}\BibitemShut {NoStop}%
\bibitem [{\citenamefont {Schrag}\ and\ \citenamefont {Xiao}(2003)}]{schrag03}%
  \BibitemOpen
  \bibfield  {author} {\bibinfo {author} {\bibfnamefont {B.~D.}\ \bibnamefont
  {Schrag}}\ and\ \bibinfo {author} {\bibfnamefont {G.}~\bibnamefont {Xiao}},\
  }\href {\doibase 10.1063/1.1570499} {\bibfield  {journal} {\bibinfo
  {journal} {Appl. Phys. Lett.}\ }\textbf {\bibinfo {volume} {82}},\ \bibinfo
  {pages} {3272} (\bibinfo {year} {2003})}\BibitemShut {NoStop}%
\bibitem [{\citenamefont {Knauss}\ \emph {et~al.}(2001)\citenamefont {Knauss},
  \citenamefont {Cawthorne}, \citenamefont {Lettsome}, \citenamefont {Kelly},
  \citenamefont {Chatraphorn}, \citenamefont {Fleet}, \citenamefont
  {Wellstood},\ and\ \citenamefont {Vanderlinde}}]{knauss01}%
  \BibitemOpen
  \bibfield  {author} {\bibinfo {author} {\bibfnamefont {L.~A.}\ \bibnamefont
  {Knauss}}, \bibinfo {author} {\bibfnamefont {A.~B.}\ \bibnamefont
  {Cawthorne}}, \bibinfo {author} {\bibfnamefont {N.}~\bibnamefont {Lettsome}},
  \bibinfo {author} {\bibfnamefont {S.}~\bibnamefont {Kelly}}, \bibinfo
  {author} {\bibfnamefont {S.}~\bibnamefont {Chatraphorn}}, \bibinfo {author}
  {\bibfnamefont {E.~F.}\ \bibnamefont {Fleet}}, \bibinfo {author}
  {\bibfnamefont {F.~C.}\ \bibnamefont {Wellstood}}, \ and\ \bibinfo {author}
  {\bibfnamefont {W.~E.}\ \bibnamefont {Vanderlinde}},\ }\href@noop {}
  {\bibfield  {journal} {\bibinfo  {journal} {Microelectronics Reliability}\
  }\textbf {\bibinfo {volume} {41}},\ \bibinfo {pages} {1211} (\bibinfo {year}
  {2001})}\BibitemShut {NoStop}%
\bibitem [{\citenamefont {Nowack}\ \emph {et~al.}(2013)\citenamefont {Nowack},
  \citenamefont {Spanton}, \citenamefont {Baenninger}, \citenamefont {Konig},
  \citenamefont {Kirtley}, \citenamefont {Kalisky}, \citenamefont {Ames},
  \citenamefont {Leubner}, \citenamefont {Brune}, \citenamefont {Buhmann},
  \citenamefont {Molenkamp}, \citenamefont {Goldhaber-Gordon},\ and\
  \citenamefont {Moler}}]{nowack13}%
  \BibitemOpen
  \bibfield  {author} {\bibinfo {author} {\bibfnamefont {K.~C.}\ \bibnamefont
  {Nowack}}, \bibinfo {author} {\bibfnamefont {E.~M.}\ \bibnamefont {Spanton}},
  \bibinfo {author} {\bibfnamefont {M.}~\bibnamefont {Baenninger}}, \bibinfo
  {author} {\bibfnamefont {M.}~\bibnamefont {Konig}}, \bibinfo {author}
  {\bibfnamefont {J.~R.}\ \bibnamefont {Kirtley}}, \bibinfo {author}
  {\bibfnamefont {B.}~\bibnamefont {Kalisky}}, \bibinfo {author} {\bibfnamefont
  {C.}~\bibnamefont {Ames}}, \bibinfo {author} {\bibfnamefont {P.}~\bibnamefont
  {Leubner}}, \bibinfo {author} {\bibfnamefont {C.}~\bibnamefont {Brune}},
  \bibinfo {author} {\bibfnamefont {H.}~\bibnamefont {Buhmann}}, \bibinfo
  {author} {\bibfnamefont {L.~W.}\ \bibnamefont {Molenkamp}}, \bibinfo {author}
  {\bibfnamefont {D.}~\bibnamefont {Goldhaber-Gordon}}, \ and\ \bibinfo
  {author} {\bibfnamefont {K.~A.}\ \bibnamefont {Moler}},\ }\href {\doibase
  10.1038/nmat3682} {\bibfield  {journal} {\bibinfo  {journal} {Nat. Mater.}\
  }\textbf {\bibinfo {volume} {12}},\ \bibinfo {pages} {787} (\bibinfo {year}
  {2013})}\BibitemShut {NoStop}%
\bibitem [{\citenamefont {Shibata}\ \emph {et~al.}(2015)\citenamefont
  {Shibata}, \citenamefont {Nomura}, \citenamefont {Kashiwaya}, \citenamefont
  {Kashiwaya}, \citenamefont {Ishiguro},\ and\ \citenamefont
  {Takayanagi}}]{shibata15}%
  \BibitemOpen
  \bibfield  {author} {\bibinfo {author} {\bibfnamefont {Y.}~\bibnamefont
  {Shibata}}, \bibinfo {author} {\bibfnamefont {S.}~\bibnamefont {Nomura}},
  \bibinfo {author} {\bibfnamefont {H.}~\bibnamefont {Kashiwaya}}, \bibinfo
  {author} {\bibfnamefont {S.}~\bibnamefont {Kashiwaya}}, \bibinfo {author}
  {\bibfnamefont {R.}~\bibnamefont {Ishiguro}}, \ and\ \bibinfo {author}
  {\bibfnamefont {H.}~\bibnamefont {Takayanagi}},\ }\href {\doibase Article}
  {\bibfield  {journal} {\bibinfo  {journal} {Sci. Rep.}\ }\textbf {\bibinfo
  {volume} {5}},\ \bibinfo {pages} {15097} (\bibinfo {year}
  {2015})}\BibitemShut {NoStop}%
\bibitem [{\citenamefont {Steinert}\ \emph {et~al.}(2010)\citenamefont
  {Steinert}, \citenamefont {Dolde}, \citenamefont {Neumann}, \citenamefont
  {Aird}, \citenamefont {Naydenov}, \citenamefont {Balasubramanian},
  \citenamefont {Jelezko},\ and\ \citenamefont {Wrachtrup}}]{steinert10}%
  \BibitemOpen
  \bibfield  {author} {\bibinfo {author} {\bibfnamefont {S.}~\bibnamefont
  {Steinert}}, \bibinfo {author} {\bibfnamefont {F.}~\bibnamefont {Dolde}},
  \bibinfo {author} {\bibfnamefont {P.}~\bibnamefont {Neumann}}, \bibinfo
  {author} {\bibfnamefont {A.}~\bibnamefont {Aird}}, \bibinfo {author}
  {\bibfnamefont {B.}~\bibnamefont {Naydenov}}, \bibinfo {author}
  {\bibfnamefont {G.}~\bibnamefont {Balasubramanian}}, \bibinfo {author}
  {\bibfnamefont {F.}~\bibnamefont {Jelezko}}, \ and\ \bibinfo {author}
  {\bibfnamefont {J.}~\bibnamefont {Wrachtrup}},\ }\href {\doibase
  10.1063/1.3385689} {\bibfield  {journal} {\bibinfo  {journal} {Rev. Sci.
  Instrum.}\ }\textbf {\bibinfo {volume} {81}},\ \bibinfo {pages} {43705}
  (\bibinfo {year} {2010})}\BibitemShut {NoStop}%
\bibitem [{\citenamefont {Chipaux}\ \emph {et~al.}(2015)\citenamefont
  {Chipaux}, \citenamefont {Tallaire}, \citenamefont {Achard}, \citenamefont
  {Pezzagna}, \citenamefont {Meijer}, \citenamefont {Jacques}, \citenamefont
  {Roch},\ and\ \citenamefont {Debuisschert}}]{chipaux15}%
  \BibitemOpen
  \bibfield  {author} {\bibinfo {author} {\bibfnamefont {M.}~\bibnamefont
  {Chipaux}}, \bibinfo {author} {\bibfnamefont {A.}~\bibnamefont {Tallaire}},
  \bibinfo {author} {\bibfnamefont {J.}~\bibnamefont {Achard}}, \bibinfo
  {author} {\bibfnamefont {S.}~\bibnamefont {Pezzagna}}, \bibinfo {author}
  {\bibfnamefont {J.}~\bibnamefont {Meijer}}, \bibinfo {author} {\bibfnamefont
  {V.}~\bibnamefont {Jacques}}, \bibinfo {author} {\bibfnamefont {J.~F.}\
  \bibnamefont {Roch}}, \ and\ \bibinfo {author} {\bibfnamefont
  {T.}~\bibnamefont {Debuisschert}},\ }\href {\doibase
  10.1140/epjd/e2015-60080-1} {\bibfield  {journal} {\bibinfo  {journal} {Eur.
  Phys. J. D}\ }\textbf {\bibinfo {volume} {69}},\ \bibinfo {pages} {166}
  (\bibinfo {year} {2015})}\BibitemShut {NoStop}%
\bibitem [{\citenamefont {Nowodzinski}\ \emph {et~al.}(2015)\citenamefont
  {Nowodzinski}, \citenamefont {Chipaux}, \citenamefont {Toraille},
  \citenamefont {Jacques}, \citenamefont {Roch},\ and\ \citenamefont
  {Debuisschert}}]{nowodzinski15}%
  \BibitemOpen
  \bibfield  {author} {\bibinfo {author} {\bibfnamefont {A.}~\bibnamefont
  {Nowodzinski}}, \bibinfo {author} {\bibfnamefont {M.}~\bibnamefont
  {Chipaux}}, \bibinfo {author} {\bibfnamefont {L.}~\bibnamefont {Toraille}},
  \bibinfo {author} {\bibfnamefont {V.}~\bibnamefont {Jacques}}, \bibinfo
  {author} {\bibfnamefont {J.~F.}\ \bibnamefont {Roch}}, \ and\ \bibinfo
  {author} {\bibfnamefont {T.}~\bibnamefont {Debuisschert}},\ }\href {\doibase
  10.1016/j.microrel.2015.06.069} {\bibfield  {journal} {\bibinfo  {journal}
  {Microelectronics Reliability}\ }\textbf {\bibinfo {volume} {55}},\ \bibinfo
  {pages} {1549} (\bibinfo {year} {2015})}\BibitemShut {NoStop}%
\bibitem [{\citenamefont {Tetienne}\ \emph {et~al.}(2016)\citenamefont
  {Tetienne}, \citenamefont {Dontschuk}, \citenamefont {Broadway},
  \citenamefont {Stacey}, \citenamefont {Simpson},\ and\ \citenamefont
  {Hollenberg}}]{tetienne16}%
  \BibitemOpen
  \bibfield  {author} {\bibinfo {author} {\bibfnamefont {J.~P.}\ \bibnamefont
  {Tetienne}}, \bibinfo {author} {\bibfnamefont {N.}~\bibnamefont {Dontschuk}},
  \bibinfo {author} {\bibfnamefont {D.~A.}\ \bibnamefont {Broadway}}, \bibinfo
  {author} {\bibfnamefont {A.}~\bibnamefont {Stacey}}, \bibinfo {author}
  {\bibfnamefont {D.~A.}\ \bibnamefont {Simpson}}, \ and\ \bibinfo {author}
  {\bibfnamefont {L.~C.~L.}\ \bibnamefont {Hollenberg}},\ }\href
  {https://arxiv.org/abs/1609.09208} {\bibfield  {journal} {\bibinfo  {journal}
  {arXiv:1609.09208}\ } (\bibinfo {year} {2016})}\BibitemShut {NoStop}%
\bibitem [{\citenamefont {Topinka}\ \emph {et~al.}(2000)\citenamefont
  {Topinka}, \citenamefont {Leroy}, \citenamefont {Shaw}, \citenamefont
  {Heller}, \citenamefont {Westervelt}, \citenamefont {Maranowski},\ and\
  \citenamefont {Gossard}}]{topinka00}%
  \BibitemOpen
  \bibfield  {author} {\bibinfo {author} {\bibfnamefont {M.~A.}\ \bibnamefont
  {Topinka}}, \bibinfo {author} {\bibfnamefont {B.~J.}\ \bibnamefont {Leroy}},
  \bibinfo {author} {\bibfnamefont {S.~E.~J.}\ \bibnamefont {Shaw}}, \bibinfo
  {author} {\bibfnamefont {E.~J.}\ \bibnamefont {Heller}}, \bibinfo {author}
  {\bibfnamefont {R.~M.}\ \bibnamefont {Westervelt}}, \bibinfo {author}
  {\bibfnamefont {K.~D.}\ \bibnamefont {Maranowski}}, \ and\ \bibinfo {author}
  {\bibfnamefont {A.~C.}\ \bibnamefont {Gossard}},\ }\href {\doibase
  10.1126/science.289.5488.2323} {\bibfield  {journal} {\bibinfo  {journal}
  {Science}\ }\textbf {\bibinfo {volume} {289}},\ \bibinfo {pages} {2323}
  (\bibinfo {year} {2000})}\BibitemShut {NoStop}%
\bibitem [{\citenamefont {Topinka}\ \emph {et~al.}(2001)\citenamefont
  {Topinka}, \citenamefont {Leroy}, \citenamefont {Westervelt}, \citenamefont
  {Shaw}, \citenamefont {Fleischmann}, \citenamefont {Heller}, \citenamefont
  {Maranowski},\ and\ \citenamefont {Gossard}}]{topinka01}%
  \BibitemOpen
  \bibfield  {author} {\bibinfo {author} {\bibfnamefont {M.~A.}\ \bibnamefont
  {Topinka}}, \bibinfo {author} {\bibfnamefont {B.~J.}\ \bibnamefont {Leroy}},
  \bibinfo {author} {\bibfnamefont {R.~M.}\ \bibnamefont {Westervelt}},
  \bibinfo {author} {\bibfnamefont {S.~E.~J.}\ \bibnamefont {Shaw}}, \bibinfo
  {author} {\bibfnamefont {R.}~\bibnamefont {Fleischmann}}, \bibinfo {author}
  {\bibfnamefont {E.~J.}\ \bibnamefont {Heller}}, \bibinfo {author}
  {\bibfnamefont {K.~D.}\ \bibnamefont {Maranowski}}, \ and\ \bibinfo {author}
  {\bibfnamefont {A.~C.}\ \bibnamefont {Gossard}},\ }\href {\doibase
  10.1038/35065553} {\bibfield  {journal} {\bibinfo  {journal} {Nature}\
  }\textbf {\bibinfo {volume} {410}},\ \bibinfo {pages} {183} (\bibinfo {year}
  {2001})}\BibitemShut {NoStop}%
\bibitem [{\citenamefont {Ihn}(2010)}]{ihn10}%
  \BibitemOpen
  \bibfield  {author} {\bibinfo {author} {\bibfnamefont {T.}~\bibnamefont
  {Ihn}},\ }\href@noop {} {\bibfield  {journal} {\bibinfo  {journal} {Oxford
  University Press, USA}\ } (\bibinfo {year} {2010})}\BibitemShut {NoStop}%
\bibitem [{\citenamefont {Son}\ \emph {et~al.}(2006)\citenamefont {Son},
  \citenamefont {Cohen},\ and\ \citenamefont {Louie}}]{son06}%
  \BibitemOpen
  \bibfield  {author} {\bibinfo {author} {\bibfnamefont {Y.}~\bibnamefont
  {Son}}, \bibinfo {author} {\bibfnamefont {M.~L.}\ \bibnamefont {Cohen}}, \
  and\ \bibinfo {author} {\bibfnamefont {S.~G.}\ \bibnamefont {Louie}},\ }\href
  {\doibase 10.1038/nature05180} {\bibfield  {journal} {\bibinfo  {journal}
  {Nature}\ }\textbf {\bibinfo {volume} {444}},\ \bibinfo {pages} {347}
  (\bibinfo {year} {2006})}\BibitemShut {NoStop}%
\bibitem [{\citenamefont {Zarbo}\ and\ \citenamefont
  {Nikolic}(2007)}]{zarbo07}%
  \BibitemOpen
  \bibfield  {author} {\bibinfo {author} {\bibfnamefont {L.~P.}\ \bibnamefont
  {Zarbo}}\ and\ \bibinfo {author} {\bibfnamefont {B.~K.}\ \bibnamefont
  {Nikolic}},\ }\href {\doibase 10.1209/0295-5075/80/47001} {\bibfield
  {journal} {\bibinfo  {journal} {Europhysics Letters}\ }\textbf {\bibinfo
  {volume} {80}},\ \bibinfo {pages} {47001} (\bibinfo {year}
  {2007})}\BibitemShut {NoStop}%
\bibitem [{\citenamefont {Bleszynski-Jayich}\ \emph {et~al.}(2009)\citenamefont
  {Bleszynski-Jayich}, \citenamefont {Shanks}, \citenamefont {Peaudecerf},
  \citenamefont {Ginossar}, \citenamefont {von Oppen}, \citenamefont
  {Glazman},\ and\ \citenamefont {Harris}}]{bleszynski09}%
  \BibitemOpen
  \bibfield  {author} {\bibinfo {author} {\bibfnamefont {A.~C.}\ \bibnamefont
  {Bleszynski-Jayich}}, \bibinfo {author} {\bibfnamefont {W.~E.}\ \bibnamefont
  {Shanks}}, \bibinfo {author} {\bibfnamefont {B.}~\bibnamefont {Peaudecerf}},
  \bibinfo {author} {\bibfnamefont {E.}~\bibnamefont {Ginossar}}, \bibinfo
  {author} {\bibfnamefont {F.}~\bibnamefont {von Oppen}}, \bibinfo {author}
  {\bibfnamefont {L.}~\bibnamefont {Glazman}}, \ and\ \bibinfo {author}
  {\bibfnamefont {J.~G.~E.}\ \bibnamefont {Harris}},\ }\href {\doibase
  10.1126/science.1178139} {\bibfield  {journal} {\bibinfo  {journal}
  {Science}\ }\textbf {\bibinfo {volume} {326}},\ \bibinfo {pages} {272}
  (\bibinfo {year} {2009})}\BibitemShut {NoStop}%
\bibitem [{\citenamefont {Bluhm}\ \emph {et~al.}(2009)\citenamefont {Bluhm},
  \citenamefont {Koshnick}, \citenamefont {Bert}, \citenamefont {Huber},\ and\
  \citenamefont {Moler}}]{bluhm09current}%
  \BibitemOpen
  \bibfield  {author} {\bibinfo {author} {\bibfnamefont {H.}~\bibnamefont
  {Bluhm}}, \bibinfo {author} {\bibfnamefont {N.~C.}\ \bibnamefont {Koshnick}},
  \bibinfo {author} {\bibfnamefont {J.~A.}\ \bibnamefont {Bert}}, \bibinfo
  {author} {\bibfnamefont {M.~E.}\ \bibnamefont {Huber}}, \ and\ \bibinfo
  {author} {\bibfnamefont {K.~A.}\ \bibnamefont {Moler}},\ }\href {\doibase
  10.1103/PhysRevLett.102.136802} {\bibfield  {journal} {\bibinfo  {journal}
  {Physical Review Letters}\ }\textbf {\bibinfo {volume} {102}},\ \bibinfo
  {pages} {136802} (\bibinfo {year} {2009})}\BibitemShut {NoStop}%
\bibitem [{\citenamefont {Friedel}(1952)}]{friedel52}%
  \BibitemOpen
  \bibfield  {author} {\bibinfo {author} {\bibfnamefont {J.}~\bibnamefont
  {Friedel}},\ }\href {\doibase 10.1080/14786440208561086} {\bibfield
  {journal} {\bibinfo  {journal} {Philos. Mag.}\ }\textbf {\bibinfo {volume}
  {43}},\ \bibinfo {pages} {153} (\bibinfo {year} {1952})}\BibitemShut
  {NoStop}%
\bibitem [{\citenamefont {Cheianov}\ and\ \citenamefont
  {Fal'ko}(2006)}]{cheianov06}%
  \BibitemOpen
  \bibfield  {author} {\bibinfo {author} {\bibfnamefont {V.}~\bibnamefont
  {Cheianov}}\ and\ \bibinfo {author} {\bibfnamefont {V.}~\bibnamefont
  {Fal'ko}},\ }\href {\doibase 10.1103/PhysRevLett.97.226801} {\bibfield
  {journal} {\bibinfo  {journal} {Phys. Rev. Lett.}\ }\textbf {\bibinfo
  {volume} {97}},\ \bibinfo {pages} {226801} (\bibinfo {year}
  {2006})}\BibitemShut {NoStop}%
\bibitem [{\citenamefont {Anahory}\ \emph {et~al.}(2014)\citenamefont
  {Anahory}, \citenamefont {Reiner}, \citenamefont {Embon}, \citenamefont
  {Halbertal}, \citenamefont {Yakovenko}, \citenamefont {Myasoedov},
  \citenamefont {Rappaport}, \citenamefont {Huber},\ and\ \citenamefont
  {Zeldov}}]{anahory14}%
  \BibitemOpen
  \bibfield  {author} {\bibinfo {author} {\bibfnamefont {Y.}~\bibnamefont
  {Anahory}}, \bibinfo {author} {\bibfnamefont {J.}~\bibnamefont {Reiner}},
  \bibinfo {author} {\bibfnamefont {L.}~\bibnamefont {Embon}}, \bibinfo
  {author} {\bibfnamefont {D.}~\bibnamefont {Halbertal}}, \bibinfo {author}
  {\bibfnamefont {A.}~\bibnamefont {Yakovenko}}, \bibinfo {author}
  {\bibfnamefont {Y.}~\bibnamefont {Myasoedov}}, \bibinfo {author}
  {\bibfnamefont {M.~L.}\ \bibnamefont {Rappaport}}, \bibinfo {author}
  {\bibfnamefont {M.~E.}\ \bibnamefont {Huber}}, \ and\ \bibinfo {author}
  {\bibfnamefont {E.}~\bibnamefont {Zeldov}},\ }\href {\doibase
  10.1021/nl503022q} {\bibfield  {journal} {\bibinfo  {journal} {Nano Lett.}\
  }\textbf {\bibinfo {volume} {14}},\ \bibinfo {pages} {6481} (\bibinfo {year}
  {2014})}\BibitemShut {NoStop}%
\bibitem [{\citenamefont {Zeldov}()}]{zeldov16}%
  \BibitemOpen
  \bibfield  {author} {\bibinfo {author} {\bibfnamefont {E.}~\bibnamefont
  {Zeldov}},\ }\href
  {https://www.ethz.ch/content/specialinterest/phys/physics-colloquium/en/programme/previous/spring16/zeldov.html}
  {\bibinfo  {journal} {Presented at the Zurich Physics Colloquium, April 20,
  2016}\ }\BibitemShut {NoStop}%
\bibitem [{\citenamefont {Degen}(2008)}]{degen08apl}%
  \BibitemOpen
\bibfield  {journal} {  }\bibfield  {author} {\bibinfo {author} {\bibfnamefont
  {C.~L.}\ \bibnamefont {Degen}},\ }\href {\doibase 10.1063/1.2943282}
  {\bibfield  {journal} {\bibinfo  {journal} {Appl. Phys. Lett.}\ }\textbf
  {\bibinfo {volume} {92}},\ \bibinfo {eid} {243111} (\bibinfo {year}
  {2008})}\BibitemShut {NoStop}%
\bibitem [{\citenamefont {Gruber}\ \emph {et~al.}(1997)\citenamefont {Gruber},
  \citenamefont {Drabenstedt}, \citenamefont {Tietz}, \citenamefont {Fleury},
  \citenamefont {Wrachtrup},\ and\ \citenamefont {von
  Borczyskowski}}]{gruber97}%
  \BibitemOpen
  \bibfield  {author} {\bibinfo {author} {\bibfnamefont {A.}~\bibnamefont
  {Gruber}}, \bibinfo {author} {\bibfnamefont {A.}~\bibnamefont {Drabenstedt}},
  \bibinfo {author} {\bibfnamefont {C.}~\bibnamefont {Tietz}}, \bibinfo
  {author} {\bibfnamefont {L.}~\bibnamefont {Fleury}}, \bibinfo {author}
  {\bibfnamefont {J.}~\bibnamefont {Wrachtrup}}, \ and\ \bibinfo {author}
  {\bibfnamefont {C.}~\bibnamefont {von Borczyskowski}},\ }\href {\doibase
  10.1126/science.276.5321.2012} {\bibfield  {journal} {\bibinfo  {journal}
  {Science}\ }\textbf {\bibinfo {volume} {276}},\ \bibinfo {eid} {2012}
  (\bibinfo {year} {1997})}\BibitemShut {NoStop}%
\bibitem [{\citenamefont {Jelezko}\ and\ \citenamefont
  {Wrachtrup}(2006)}]{jelezko06}%
  \BibitemOpen
  \bibfield  {author} {\bibinfo {author} {\bibfnamefont {F.}~\bibnamefont
  {Jelezko}}\ and\ \bibinfo {author} {\bibfnamefont {J.}~\bibnamefont
  {Wrachtrup}},\ }\href {\doibase 10.1002/pssa.200671403} {\bibfield  {journal}
  {\bibinfo  {journal} {phys. stat. sol. (a)}\ }\textbf {\bibinfo {volume}
  {203}},\ \bibinfo {eid} {3207} (\bibinfo {year} {2006})}\BibitemShut
  {NoStop}%
\bibitem [{\citenamefont {Balasubramanian}\ \emph {et~al.}(2008)\citenamefont
  {Balasubramanian}, \citenamefont {Chan}, \citenamefont {Kolesov},
  \citenamefont {Al-Hmoud}, \citenamefont {Tisler}, \citenamefont {Shin},
  \citenamefont {Kim}, \citenamefont {Wojcik}, \citenamefont {Hemmer},
  \citenamefont {Krueger}, \citenamefont {Hanke}, \citenamefont
  {Leitenstorfer}, \citenamefont {Bratschitsch}, \citenamefont {Jelezko},\ and\
  \citenamefont {Wrachtrup}}]{balasubramanian08}%
  \BibitemOpen
  \bibfield  {author} {\bibinfo {author} {\bibfnamefont {G.}~\bibnamefont
  {Balasubramanian}}, \bibinfo {author} {\bibfnamefont {I.~Y.}\ \bibnamefont
  {Chan}}, \bibinfo {author} {\bibfnamefont {R.}~\bibnamefont {Kolesov}},
  \bibinfo {author} {\bibfnamefont {M.}~\bibnamefont {Al-Hmoud}}, \bibinfo
  {author} {\bibfnamefont {J.}~\bibnamefont {Tisler}}, \bibinfo {author}
  {\bibfnamefont {C.}~\bibnamefont {Shin}}, \bibinfo {author} {\bibfnamefont
  {C.}~\bibnamefont {Kim}}, \bibinfo {author} {\bibfnamefont {A.}~\bibnamefont
  {Wojcik}}, \bibinfo {author} {\bibfnamefont {P.~R.}\ \bibnamefont {Hemmer}},
  \bibinfo {author} {\bibfnamefont {A.}~\bibnamefont {Krueger}}, \bibinfo
  {author} {\bibfnamefont {T.}~\bibnamefont {Hanke}}, \bibinfo {author}
  {\bibfnamefont {A.}~\bibnamefont {Leitenstorfer}}, \bibinfo {author}
  {\bibfnamefont {R.}~\bibnamefont {Bratschitsch}}, \bibinfo {author}
  {\bibfnamefont {F.}~\bibnamefont {Jelezko}}, \ and\ \bibinfo {author}
  {\bibfnamefont {J.}~\bibnamefont {Wrachtrup}},\ }\href {\doibase
  10.1038/nature07278} {\bibfield  {journal} {\bibinfo  {journal} {Nature}\
  }\textbf {\bibinfo {volume} {455}},\ \bibinfo {eid} {648} (\bibinfo {year}
  {2008})}\BibitemShut {NoStop}%
\bibitem [{\citenamefont {Grinolds}\ \emph {et~al.}(2013)\citenamefont
  {Grinolds}, \citenamefont {Hong}, \citenamefont {Maletinsky}, \citenamefont
  {Luan}, \citenamefont {Lukin}, \citenamefont {Walsworth},\ and\ \citenamefont
  {Yacoby}}]{grinolds13}%
  \BibitemOpen
  \bibfield  {author} {\bibinfo {author} {\bibfnamefont {M.~S.}\ \bibnamefont
  {Grinolds}}, \bibinfo {author} {\bibfnamefont {S.}~\bibnamefont {Hong}},
  \bibinfo {author} {\bibfnamefont {P.}~\bibnamefont {Maletinsky}}, \bibinfo
  {author} {\bibfnamefont {L.}~\bibnamefont {Luan}}, \bibinfo {author}
  {\bibfnamefont {M.~.~D.}\ \bibnamefont {Lukin}}, \bibinfo {author}
  {\bibfnamefont {R.~L.}\ \bibnamefont {Walsworth}}, \ and\ \bibinfo {author}
  {\bibfnamefont {A.}~\bibnamefont {Yacoby}},\ }\href {\doibase
  10.1038/nphys2543} {\bibfield  {journal} {\bibinfo  {journal} {Nat. Phys.}\
  }\textbf {\bibinfo {volume} {9}},\ \bibinfo {pages} {215} (\bibinfo {year}
  {2013})}\BibitemShut {NoStop}%
\bibitem [{\citenamefont {Rondin}\ \emph {et~al.}(2012)\citenamefont {Rondin},
  \citenamefont {Tetienne}, \citenamefont {Spinicelli}, \citenamefont {dal
  Savio}, \citenamefont {Karrai}, \citenamefont {Dantelle}, \citenamefont
  {Thiaville}, \citenamefont {Rohart}, \citenamefont {Roch},\ and\
  \citenamefont {Jacques}}]{rondin12}%
  \BibitemOpen
  \bibfield  {author} {\bibinfo {author} {\bibfnamefont {L.}~\bibnamefont
  {Rondin}}, \bibinfo {author} {\bibfnamefont {J.~P.}\ \bibnamefont
  {Tetienne}}, \bibinfo {author} {\bibfnamefont {P.}~\bibnamefont
  {Spinicelli}}, \bibinfo {author} {\bibfnamefont {C.}~\bibnamefont {dal
  Savio}}, \bibinfo {author} {\bibfnamefont {K.}~\bibnamefont {Karrai}},
  \bibinfo {author} {\bibfnamefont {G.}~\bibnamefont {Dantelle}}, \bibinfo
  {author} {\bibfnamefont {A.}~\bibnamefont {Thiaville}}, \bibinfo {author}
  {\bibfnamefont {S.}~\bibnamefont {Rohart}}, \bibinfo {author} {\bibfnamefont
  {J.~F.}\ \bibnamefont {Roch}}, \ and\ \bibinfo {author} {\bibfnamefont
  {V.}~\bibnamefont {Jacques}},\ }\href {\doibase 10.1063/1.3703128} {\bibfield
   {journal} {\bibinfo  {journal} {Appl. Phys. Lett.}\ }\textbf {\bibinfo
  {volume} {100}},\ \bibinfo {pages} {153118} (\bibinfo {year}
  {2012})}\BibitemShut {NoStop}%
\bibitem [{\citenamefont {Maletinsky}\ \emph {et~al.}(2012)\citenamefont
  {Maletinsky}, \citenamefont {Hong}, \citenamefont {Grinolds}, \citenamefont
  {Hausmann}, \citenamefont {Lukin}, \citenamefont {Walsworth}, \citenamefont
  {Loncar},\ and\ \citenamefont {Yacoby}}]{maletinsky12}%
  \BibitemOpen
  \bibfield  {author} {\bibinfo {author} {\bibfnamefont {P.}~\bibnamefont
  {Maletinsky}}, \bibinfo {author} {\bibfnamefont {S.}~\bibnamefont {Hong}},
  \bibinfo {author} {\bibfnamefont {M.~S.}\ \bibnamefont {Grinolds}}, \bibinfo
  {author} {\bibfnamefont {B.}~\bibnamefont {Hausmann}}, \bibinfo {author}
  {\bibfnamefont {M.~D.}\ \bibnamefont {Lukin}}, \bibinfo {author}
  {\bibfnamefont {R.~L.}\ \bibnamefont {Walsworth}}, \bibinfo {author}
  {\bibfnamefont {M.}~\bibnamefont {Loncar}}, \ and\ \bibinfo {author}
  {\bibfnamefont {A.}~\bibnamefont {Yacoby}},\ }\href {\doibase
  10.1038/NNANO.2012.50} {\bibfield  {journal} {\bibinfo  {journal} {Nat.
  Nanotechnol.}\ }\textbf {\bibinfo {volume} {7}},\ \bibinfo {pages} {320}
  (\bibinfo {year} {2012})}\BibitemShut {NoStop}%
\bibitem [{\citenamefont {Rondin}\ \emph {et~al.}(2013)\citenamefont {Rondin},
  \citenamefont {Tetienne}, \citenamefont {Rohart}, \citenamefont {Thiaville},
  \citenamefont {Hingant}, \citenamefont {Spinicelli}, \citenamefont {Roch},\
  and\ \citenamefont {Jacques}}]{rondin13}%
  \BibitemOpen
  \bibfield  {author} {\bibinfo {author} {\bibfnamefont {L.}~\bibnamefont
  {Rondin}}, \bibinfo {author} {\bibfnamefont {J.~P.}\ \bibnamefont
  {Tetienne}}, \bibinfo {author} {\bibfnamefont {S.}~\bibnamefont {Rohart}},
  \bibinfo {author} {\bibfnamefont {A.}~\bibnamefont {Thiaville}}, \bibinfo
  {author} {\bibfnamefont {T.}~\bibnamefont {Hingant}}, \bibinfo {author}
  {\bibfnamefont {P.}~\bibnamefont {Spinicelli}}, \bibinfo {author}
  {\bibfnamefont {J.~F.}\ \bibnamefont {Roch}}, \ and\ \bibinfo {author}
  {\bibfnamefont {V.}~\bibnamefont {Jacques}},\ }\href {\doibase
  10.1038/ncomms3279} {\bibfield  {journal} {\bibinfo  {journal} {Nat. Comms.}\
  }\textbf {\bibinfo {volume} {4}},\ \bibinfo {pages} {2279} (\bibinfo {year}
  {2013})}\BibitemShut {NoStop}%
\bibitem [{\citenamefont {Tetienne}\ \emph {et~al.}(2015)\citenamefont
  {Tetienne}, \citenamefont {Hingant}, \citenamefont {Martinez}, \citenamefont
  {Rohart}, \citenamefont {Thiaville}, \citenamefont {Diez}, \citenamefont
  {Garcia}, \citenamefont {Adam}, \citenamefont {Kim}, \citenamefont {Roch},
  \citenamefont {Miron}, \citenamefont {Gaudin}, \citenamefont {Vila},
  \citenamefont {Ocker}, \citenamefont {Ravelosona},\ and\ \citenamefont
  {Jacques}}]{tetienne15}%
  \BibitemOpen
  \bibfield  {author} {\bibinfo {author} {\bibfnamefont {J.~P.}\ \bibnamefont
  {Tetienne}}, \bibinfo {author} {\bibfnamefont {T.}~\bibnamefont {Hingant}},
  \bibinfo {author} {\bibfnamefont {L.~J.}\ \bibnamefont {Martinez}}, \bibinfo
  {author} {\bibfnamefont {S.}~\bibnamefont {Rohart}}, \bibinfo {author}
  {\bibfnamefont {A.}~\bibnamefont {Thiaville}}, \bibinfo {author}
  {\bibfnamefont {L.~H.}\ \bibnamefont {Diez}}, \bibinfo {author}
  {\bibfnamefont {K.}~\bibnamefont {Garcia}}, \bibinfo {author} {\bibfnamefont
  {J.~P.}\ \bibnamefont {Adam}}, \bibinfo {author} {\bibfnamefont {J.~V.}\
  \bibnamefont {Kim}}, \bibinfo {author} {\bibfnamefont {J.~F.}\ \bibnamefont
  {Roch}}, \bibinfo {author} {\bibfnamefont {I.~M.}\ \bibnamefont {Miron}},
  \bibinfo {author} {\bibfnamefont {G.}~\bibnamefont {Gaudin}}, \bibinfo
  {author} {\bibfnamefont {L.}~\bibnamefont {Vila}}, \bibinfo {author}
  {\bibfnamefont {B.}~\bibnamefont {Ocker}}, \bibinfo {author} {\bibfnamefont
  {D.}~\bibnamefont {Ravelosona}}, \ and\ \bibinfo {author} {\bibfnamefont
  {V.}~\bibnamefont {Jacques}},\ }\href {\doibase 10.1038/ncomms7733}
  {\bibfield  {journal} {\bibinfo  {journal} {Nat. Commun.}\ }\textbf {\bibinfo
  {volume} {6}} (\bibinfo {year} {2015}),\ 10.1038/ncomms7733}\BibitemShut
  {NoStop}%
\bibitem [{\citenamefont {Pelliccione}\ \emph {et~al.}(2016)\citenamefont
  {Pelliccione}, \citenamefont {Jenkins}, \citenamefont {Ovartchaiyapong},
  \citenamefont {Reetz}, \citenamefont {Emmanouilidou}, \citenamefont {Ni},\
  and\ \citenamefont {Jayich}}]{pelliccione16}%
  \BibitemOpen
  \bibfield  {author} {\bibinfo {author} {\bibfnamefont {M.}~\bibnamefont
  {Pelliccione}}, \bibinfo {author} {\bibfnamefont {A.}~\bibnamefont
  {Jenkins}}, \bibinfo {author} {\bibfnamefont {P.}~\bibnamefont
  {Ovartchaiyapong}}, \bibinfo {author} {\bibfnamefont {C.}~\bibnamefont
  {Reetz}}, \bibinfo {author} {\bibfnamefont {E.}~\bibnamefont
  {Emmanouilidou}}, \bibinfo {author} {\bibfnamefont {N.}~\bibnamefont {Ni}}, \
  and\ \bibinfo {author} {\bibfnamefont {A.~C.~B.}\ \bibnamefont {Jayich}},\
  }\href {\doibase 10.1038/NNANO.2016.68} {\bibfield  {journal} {\bibinfo
  {journal} {Nature Nanotechnology}\ }\textbf {\bibinfo {volume} {11}},\
  \bibinfo {pages} {700} (\bibinfo {year} {2016})}\BibitemShut {NoStop}%
\bibitem [{\citenamefont {Thiel}\ \emph {et~al.}(2016)\citenamefont {Thiel},
  \citenamefont {Rohner}, \citenamefont {Ganzhorn}, \citenamefont {Appel},
  \citenamefont {Neu}, \citenamefont {Muller}, \citenamefont {Kleiner},
  \citenamefont {Koelle},\ and\ \citenamefont {Maletinsky}}]{thiel16}%
  \BibitemOpen
  \bibfield  {author} {\bibinfo {author} {\bibfnamefont {L.}~\bibnamefont
  {Thiel}}, \bibinfo {author} {\bibfnamefont {D.}~\bibnamefont {Rohner}},
  \bibinfo {author} {\bibfnamefont {M.}~\bibnamefont {Ganzhorn}}, \bibinfo
  {author} {\bibfnamefont {P.}~\bibnamefont {Appel}}, \bibinfo {author}
  {\bibfnamefont {E.}~\bibnamefont {Neu}}, \bibinfo {author} {\bibfnamefont
  {B.}~\bibnamefont {Muller}}, \bibinfo {author} {\bibfnamefont
  {R.}~\bibnamefont {Kleiner}}, \bibinfo {author} {\bibfnamefont
  {D.}~\bibnamefont {Koelle}}, \ and\ \bibinfo {author} {\bibfnamefont
  {P.}~\bibnamefont {Maletinsky}},\ }\href {\doibase 10.1038/NNANO.2016.63}
  {\bibfield  {journal} {\bibinfo  {journal} {Nature Nanotechnology}\ }\textbf
  {\bibinfo {volume} {11}},\ \bibinfo {pages} {677} (\bibinfo {year}
  {2016})}\BibitemShut {NoStop}%
\bibitem [{\citenamefont {K\"uhn}\ \emph {et~al.}(2001)\citenamefont {K\"uhn},
  \citenamefont {Hettich}, \citenamefont {Schmitt}, \citenamefont {Poizat},\
  and\ \citenamefont {Sandoghdar}}]{kuhn01}%
  \BibitemOpen
  \bibfield  {author} {\bibinfo {author} {\bibfnamefont {S.}~\bibnamefont
  {K\"uhn}}, \bibinfo {author} {\bibfnamefont {C.}~\bibnamefont {Hettich}},
  \bibinfo {author} {\bibfnamefont {C.}~\bibnamefont {Schmitt}}, \bibinfo
  {author} {\bibfnamefont {J.~P.}\ \bibnamefont {Poizat}}, \ and\ \bibinfo
  {author} {\bibfnamefont {V.}~\bibnamefont {Sandoghdar}},\ }\href {\doibase
  10.1046/j.1365-2818.2001.00829.x} {\bibfield  {journal} {\bibinfo  {journal}
  {J. Microsc.}\ }\textbf {\bibinfo {volume} {202}},\ \bibinfo {eid} {2}
  (\bibinfo {year} {2001})}\BibitemShut {NoStop}%
\bibitem [{\citenamefont {Dreau}\ \emph {et~al.}(2011)\citenamefont {Dreau},
  \citenamefont {Lesik}, \citenamefont {Rondin}, \citenamefont {Spinicelli},
  \citenamefont {Arcizet}, \citenamefont {Roch},\ and\ \citenamefont
  {Jacques}}]{dreau11}%
  \BibitemOpen
  \bibfield  {author} {\bibinfo {author} {\bibfnamefont {A.}~\bibnamefont
  {Dreau}}, \bibinfo {author} {\bibfnamefont {M.}~\bibnamefont {Lesik}},
  \bibinfo {author} {\bibfnamefont {L.}~\bibnamefont {Rondin}}, \bibinfo
  {author} {\bibfnamefont {P.}~\bibnamefont {Spinicelli}}, \bibinfo {author}
  {\bibfnamefont {O.}~\bibnamefont {Arcizet}}, \bibinfo {author} {\bibfnamefont
  {J.~F.}\ \bibnamefont {Roch}}, \ and\ \bibinfo {author} {\bibfnamefont
  {V.}~\bibnamefont {Jacques}},\ }\href {\doibase 10.1103/PhysRevB.84.195204}
  {\bibfield  {journal} {\bibinfo  {journal} {Phys. Rev. B}\ }\textbf {\bibinfo
  {volume} {84}},\ \bibinfo {pages} {195204} (\bibinfo {year}
  {2011})}\BibitemShut {NoStop}%
\bibitem [{sup()}]{supplemental}%
  \BibitemOpen
  \href@noop {} {\bibinfo  {journal} {See Supplemental Material accompanying
  this manuscript}\ }\BibitemShut {NoStop}%
\bibitem [{\citenamefont {Appel}\ \emph {et~al.}(2015)\citenamefont {Appel},
  \citenamefont {Ganzhorn}, \citenamefont {Neu},\ and\ \citenamefont
  {Maletinsky}}]{appel15}%
  \BibitemOpen
\bibfield  {journal} {  }\bibfield  {author} {\bibinfo {author} {\bibfnamefont
  {P.}~\bibnamefont {Appel}}, \bibinfo {author} {\bibfnamefont
  {M.}~\bibnamefont {Ganzhorn}}, \bibinfo {author} {\bibfnamefont
  {E.}~\bibnamefont {Neu}}, \ and\ \bibinfo {author} {\bibfnamefont
  {P.}~\bibnamefont {Maletinsky}},\ }\href {\doibase
  10.1088/1367-2630/17/11/112001} {\bibfield  {journal} {\bibinfo  {journal}
  {New Journal of Physics}\ }\textbf {\bibinfo {volume} {17}},\ \bibinfo
  {pages} {112001} (\bibinfo {year} {2015})}\BibitemShut {NoStop}%
\bibitem [{\citenamefont {Sparrow}(1916)}]{sparrow16}%
  \BibitemOpen
  \bibfield  {author} {\bibinfo {author} {\bibfnamefont {C.}~\bibnamefont
  {Sparrow}},\ }\href {\doibase 10.1086/142271} {\bibfield  {journal} {\bibinfo
   {journal} {Astrophysical Journal}\ }\textbf {\bibinfo {volume} {44}},\
  \bibinfo {pages} {76} (\bibinfo {year} {1916})}\BibitemShut {NoStop}%
\bibitem [{\citenamefont {Jones}\ \emph {et~al.}(1995)\citenamefont {Jones},
  \citenamefont {Blandhawthorn},\ and\ \citenamefont {Shopbell}}]{jones95}%
  \BibitemOpen
  \bibfield  {author} {\bibinfo {author} {\bibfnamefont {A.}~\bibnamefont
  {Jones}}, \bibinfo {author} {\bibfnamefont {J.}~\bibnamefont
  {Blandhawthorn}}, \ and\ \bibinfo {author} {\bibfnamefont {P.}~\bibnamefont
  {Shopbell}},\ }\href {http://adsabs.harvard.edu/abs/1995ASPC...77..503J}
  {\bibfield  {journal} {\bibinfo  {journal} {Astronomical Data Analysis
  Software and Systems IV, ASP Conference Series, edited by: R.A. Shaw, H.E.
  Payne, and J.J.E. Hayes}\ }\textbf {\bibinfo {volume} {77}},\ \bibinfo
  {pages} {503} (\bibinfo {year} {1995})}\BibitemShut {NoStop}%
\bibitem [{\citenamefont {Youn}\ \emph {et~al.}(2013)\citenamefont {Youn},
  \citenamefont {Yazdani}, \citenamefont {Patscheider},\ and\ \citenamefont
  {Park}}]{Youn13}%
  \BibitemOpen
  \bibfield  {author} {\bibinfo {author} {\bibfnamefont {S.~K.}\ \bibnamefont
  {Youn}}, \bibinfo {author} {\bibfnamefont {N.}~\bibnamefont {Yazdani}},
  \bibinfo {author} {\bibfnamefont {J.}~\bibnamefont {Patscheider}}, \ and\
  \bibinfo {author} {\bibfnamefont {H.~G.}\ \bibnamefont {Park}},\ }\href
  {\doibase 10.1039/c2ra22392a} {\bibfield  {journal} {\bibinfo  {journal} {RSC
  Advances}\ }\textbf {\bibinfo {volume} {3}},\ \bibinfo {pages} {1434}
  (\bibinfo {year} {2013})}\BibitemShut {NoStop}%
\bibitem [{\citenamefont {Appel}\ \emph {et~al.}(2016)\citenamefont {Appel},
  \citenamefont {Neu}, \citenamefont {Ganzhorn}, \citenamefont {Barfuss},
  \citenamefont {Batzer}, \citenamefont {Gratz}, \citenamefont {Tschope},\ and\
  \citenamefont {Maletinsky}}]{appel16}%
  \BibitemOpen
  \bibfield  {author} {\bibinfo {author} {\bibfnamefont {P.}~\bibnamefont
  {Appel}}, \bibinfo {author} {\bibfnamefont {E.}~\bibnamefont {Neu}}, \bibinfo
  {author} {\bibfnamefont {M.}~\bibnamefont {Ganzhorn}}, \bibinfo {author}
  {\bibfnamefont {A.}~\bibnamefont {Barfuss}}, \bibinfo {author} {\bibfnamefont
  {M.}~\bibnamefont {Batzer}}, \bibinfo {author} {\bibfnamefont
  {M.}~\bibnamefont {Gratz}}, \bibinfo {author} {\bibfnamefont
  {A.}~\bibnamefont {Tschope}}, \ and\ \bibinfo {author} {\bibfnamefont
  {P.}~\bibnamefont {Maletinsky}},\ }\href {\doibase 10.1063/1.4952953}
  {\bibfield  {journal} {\bibinfo  {journal} {Rev. Sci. Instr.}\ }\textbf
  {\bibinfo {volume} {87}},\ \bibinfo {pages} {063703} (\bibinfo {year}
  {2016})}\BibitemShut {NoStop}%
\bibitem [{\citenamefont {Taylor}\ \emph {et~al.}(2008)\citenamefont {Taylor},
  \citenamefont {Cappellaro}, \citenamefont {Childress}, \citenamefont {Jiang},
  \citenamefont {Budker}, \citenamefont {Hemmer}, \citenamefont {A.Yacoby},
  \citenamefont {Walsworth},\ and\ \citenamefont {Lukin}}]{taylor08}%
  \BibitemOpen
  \bibfield  {author} {\bibinfo {author} {\bibfnamefont {J.~M.}\ \bibnamefont
  {Taylor}}, \bibinfo {author} {\bibfnamefont {P.}~\bibnamefont {Cappellaro}},
  \bibinfo {author} {\bibfnamefont {L.}~\bibnamefont {Childress}}, \bibinfo
  {author} {\bibfnamefont {L.}~\bibnamefont {Jiang}}, \bibinfo {author}
  {\bibfnamefont {D.}~\bibnamefont {Budker}}, \bibinfo {author} {\bibfnamefont
  {P.~R.}\ \bibnamefont {Hemmer}}, \bibinfo {author} {\bibnamefont {A.Yacoby}},
  \bibinfo {author} {\bibfnamefont {R.}~\bibnamefont {Walsworth}}, \ and\
  \bibinfo {author} {\bibfnamefont {M.~D.}\ \bibnamefont {Lukin}},\ }\href
  {\doibase 10.1038/nphys1075} {\bibfield  {journal} {\bibinfo  {journal}
  {Nature Physics}\ }\textbf {\bibinfo {volume} {4}},\ \bibinfo {eid} {810}
  (\bibinfo {year} {2008})}\BibitemShut {NoStop}%
\bibitem [{\citenamefont {Maze}\ \emph {et~al.}(2008)\citenamefont {Maze},
  \citenamefont {Stanwix}, \citenamefont {Hodges}, \citenamefont {Hong},
  \citenamefont {Taylor}, \citenamefont {Cappellaro}, \citenamefont {Jiang},
  \citenamefont {Dutt}, \citenamefont {Togan}, \citenamefont {Zibrov},
  \citenamefont {Yacoby}, \citenamefont {Walsworth},\ and\ \citenamefont
  {Lukin}}]{maze08}%
  \BibitemOpen
  \bibfield  {author} {\bibinfo {author} {\bibfnamefont {J.~R.}\ \bibnamefont
  {Maze}}, \bibinfo {author} {\bibfnamefont {P.~L.}\ \bibnamefont {Stanwix}},
  \bibinfo {author} {\bibfnamefont {J.~S.}\ \bibnamefont {Hodges}}, \bibinfo
  {author} {\bibfnamefont {S.}~\bibnamefont {Hong}}, \bibinfo {author}
  {\bibfnamefont {J.~M.}\ \bibnamefont {Taylor}}, \bibinfo {author}
  {\bibfnamefont {P.}~\bibnamefont {Cappellaro}}, \bibinfo {author}
  {\bibfnamefont {L.}~\bibnamefont {Jiang}}, \bibinfo {author} {\bibfnamefont
  {M.~V.~G.}\ \bibnamefont {Dutt}}, \bibinfo {author} {\bibfnamefont
  {E.}~\bibnamefont {Togan}}, \bibinfo {author} {\bibfnamefont {A.~S.}\
  \bibnamefont {Zibrov}}, \bibinfo {author} {\bibfnamefont {A.}~\bibnamefont
  {Yacoby}}, \bibinfo {author} {\bibfnamefont {R.~L.}\ \bibnamefont
  {Walsworth}}, \ and\ \bibinfo {author} {\bibfnamefont {M.~D.}\ \bibnamefont
  {Lukin}},\ }\href {\doibase 10.1038/nature07279} {\bibfield  {journal}
  {\bibinfo  {journal} {Nature}\ }\textbf {\bibinfo {volume} {455}},\ \bibinfo
  {eid} {644} (\bibinfo {year} {2008})}\BibitemShut {NoStop}%
\bibitem [{\citenamefont {Nusran}\ \emph {et~al.}(2012)\citenamefont {Nusran},
  \citenamefont {Momeen},\ and\ \citenamefont {Dutt}}]{nusran12}%
  \BibitemOpen
  \bibfield  {author} {\bibinfo {author} {\bibfnamefont {N.~M.}\ \bibnamefont
  {Nusran}}, \bibinfo {author} {\bibfnamefont {M.~U.}\ \bibnamefont {Momeen}},
  \ and\ \bibinfo {author} {\bibfnamefont {M.~V.~G.}\ \bibnamefont {Dutt}},\
  }\href {\doibase 10.1038/NNANO.2011.225} {\bibfield  {journal} {\bibinfo
  {journal} {Nat. Nanotechnol.}\ }\textbf {\bibinfo {volume} {7}},\ \bibinfo
  {pages} {109} (\bibinfo {year} {2012})}\BibitemShut {NoStop}%
\bibitem [{\citenamefont {Waldherr}\ \emph {et~al.}(2012)\citenamefont
  {Waldherr}, \citenamefont {Beck}, \citenamefont {Neumann}, \citenamefont
  {Said}, \citenamefont {Nitsche}, \citenamefont {Markham}, \citenamefont
  {Twitchen}, \citenamefont {Twamley}, \citenamefont {Jelezko},\ and\
  \citenamefont {Wrachtrup}}]{waldherr12}%
  \BibitemOpen
  \bibfield  {author} {\bibinfo {author} {\bibfnamefont {G.}~\bibnamefont
  {Waldherr}}, \bibinfo {author} {\bibfnamefont {J.}~\bibnamefont {Beck}},
  \bibinfo {author} {\bibfnamefont {P.}~\bibnamefont {Neumann}}, \bibinfo
  {author} {\bibfnamefont {R.~S.}\ \bibnamefont {Said}}, \bibinfo {author}
  {\bibfnamefont {M.}~\bibnamefont {Nitsche}}, \bibinfo {author} {\bibfnamefont
  {M.~L.}\ \bibnamefont {Markham}}, \bibinfo {author} {\bibfnamefont {D.~J.}\
  \bibnamefont {Twitchen}}, \bibinfo {author} {\bibfnamefont {J.}~\bibnamefont
  {Twamley}}, \bibinfo {author} {\bibfnamefont {F.}~\bibnamefont {Jelezko}}, \
  and\ \bibinfo {author} {\bibfnamefont {J.}~\bibnamefont {Wrachtrup}},\ }\href
  {\doibase 10.1038/NNANO.2011.224} {\bibfield  {journal} {\bibinfo  {journal}
  {Nat. Nanotechnol.}\ }\textbf {\bibinfo {volume} {7}},\ \bibinfo {pages}
  {105} (\bibinfo {year} {2012})}\BibitemShut {NoStop}%
\bibitem [{\citenamefont {Bonato}\ \emph {et~al.}(2016)\citenamefont {Bonato},
  \citenamefont {Blok}, \citenamefont {Dinani}, \citenamefont {Berry},
  \citenamefont {Markham}, \citenamefont {Twitchen},\ and\ \citenamefont
  {Hanson}}]{bonato16}%
  \BibitemOpen
  \bibfield  {author} {\bibinfo {author} {\bibfnamefont {C.}~\bibnamefont
  {Bonato}}, \bibinfo {author} {\bibfnamefont {M.~S.}\ \bibnamefont {Blok}},
  \bibinfo {author} {\bibfnamefont {H.~T.}\ \bibnamefont {Dinani}}, \bibinfo
  {author} {\bibfnamefont {D.~W.}\ \bibnamefont {Berry}}, \bibinfo {author}
  {\bibfnamefont {M.~L.}\ \bibnamefont {Markham}}, \bibinfo {author}
  {\bibfnamefont {D.~J.}\ \bibnamefont {Twitchen}}, \ and\ \bibinfo {author}
  {\bibfnamefont {R.}~\bibnamefont {Hanson}},\ }\href {\doibase
  10.1038/nnano.2015.261} {\bibfield  {journal} {\bibinfo  {journal} {Nat.
  Nano.}\ }\textbf {\bibinfo {volume} {11}},\ \bibinfo {pages} {247} (\bibinfo
  {year} {2016})}\BibitemShut {NoStop}%
\end{thebibliography}%







\end{document}